%% file: main.tex
\documentclass[ALICE,manyauthors]{cernphprep}
\usepackage[comma,square,numbers,sort&compress]{natbib}

\usepackage{lineno}
\usepackage{xspace}
\usepackage{hyperref}

\usepackage[separate-uncertainty=true,locale=US,table-align-uncertainty]{siunitx}
\DeclareSIUnit[quantity-product = ]\percent{\char`\%}
\usepackage{physics}

\usepackage{booktabs}
\usepackage{adjustbox}
\usepackage{rotating}

\usepackage{mathtools}

\usepackage[T1]{fontenc}
\usepackage{orcidlink}

\begin{document}
\input{commands.tex}

\begin{titlepage}
\PHyear{2025}       
\PHnumber{024}      
\PHdate{17 February}  

\title{Measurement of isolated prompt photon production \\in pp and p--Pb collisions at the LHC}
\ShortTitle{Isolated prompt photon production in pp and p--Pb collisions at the LHC}  

\Collaboration{ALICE Collaboration\thanks{See Appendix~\ref{app:collab} for the list of collaboration members}}
\ShortAuthor{ALICE Collaboration} 

\begin{abstract}
This paper presents the measurement of the isolated prompt photon inclusive production cross section in pp and p--Pb collisions by the ALICE Collaboration at the LHC.
The measurement is performed in p--Pb collisions at centre-of-mass energies per nucleon pair of $\sqrt{s_{\text{NN}}}=\SI{5.02}{TeV}$ and \SI{8.16}{TeV}, as well as in pp collisions at $\sqrt{s}=\SI{5.02}{TeV}$ and \SI{8}{TeV}.
The cross section is obtained at midrapidity $(|y|<0.7)$ using a charged-track  based isolation momentum $p_{\rm T}^{\text{iso,~ch}}<\SI{1.5}{GeV}/c$ in a cone with radius $R=0.4$.
The data for both collision systems are well reproduced by perturbative QCD (pQCD) calculations at next-to-leading order (NLO) using recent parton distribution functions for free (PDF) and bound (nPDF) nucleons.
Furthermore, the nuclear modification factor $R_{\text{pA}}$ for both collision energies is consistent with unity for \pt $>\SI{20}{GeV}/c$. 
However, deviations from unity ($R_{\rm pA}<1$) of up to \SI{20}{\percent} are observed for \pt $<\SI{20}{GeV}/c$ with limited significance, indicating the possible presence of nuclear effects in the initial state of the collision.
The suppression increases with decreasing \pt with a significance of $2.3\sigma$ for a non-zero slope and yields $R_{\rm pA}<1$ with a significance of $1.8\sigma$ at $\sqrt{s_{\rm NN}}=\SI{8.16}{TeV}$  for \pt $<\SI{20}{GeV}/c$. 
In addition, a significance of $1.1\sigma$ is observed for $R_{\rm pA}<1$ at the lower collision energy $\sqrt{s_{\rm NN}}=\SI{5.02}{TeV}$ for $\pt < \SI{14}{GeV}/c$. 
The magnitude and shape of the suppression are consistent with pQCD predictions at NLO using nPDFs that incorporate nuclear shadowing effects in the Pb nucleus.

\end{abstract}
\end{titlepage}

\setcounter{page}{2} 


\section{Introduction} 
Understanding the dynamics of partons in nuclear matter is a key goal of nuclear physics.
Measurements of inclusive particle and jet production cross sections at large transverse momenta in proton--proton (pp), proton--nucleus (pA), and nucleus--nucleus (AA) collisions are important tools to study nuclear matter, as their production occurs early in the evolution of the collision, via hard scatterings of incoming partons~\cite{journeyQCD}.
Modification of their production rates in pA and AA collisions with respect to pp collisions are attributed to the presence of nuclear matter, which can affect the initial- and final-state of the collision.
Initial state effects include changes of the parton distributions inside the nucleus (e.g. gluon shadowing~\cite{shadowing} and saturation~\cite{CGC}), isospin effects, and initial-state energy loss~\cite{initialstateenergyloss} of the incoming partons.
Final state effects reflect strong interactions of the produced particles with the hot~\cite{journeyQCD,hotnuclearmatter}  and cold~\cite{FCEL} strongly-interacting medium.
The modification of production rates is commonly quantified using 
nuclear modification factors $R_{\text{\tiny pA}}$ ($R_{\text{\tiny AA}}$), 
which is  
given by the ratio of cross sections in pA (AA) collisions 
to those in pp, 
scaled by an appropriate normalisation to account for the number of nucleon--nucleon collisions occurring in the considered pA (AA) collisions~\cite{glauber,glauber2}.

Since photons interact with other particles only through the electromagnetic interaction, their scattering with the strongly-interacting medium is rare and the photon mean free path is large.
This makes them a valuable probe for discriminating initial and final state contributions to the yield modifications observed in nuclear collisions.
Photons produced directly in the hard scatterings are referred to as prompt photons. 
They are sensitive to the gluon densities in the colliding hadrons at leading order through the quantum chromodynamics (QCD) Compton scattering process ($\rm{qg}\rightarrow \gamma \rm{q}$), which is the dominant contribution to the prompt photon production cross section at the Large Hadron Collider (LHC)~\cite{jetphox}.
Prompt photons are also produced by collinear fragmentation of an outgoing parton, which is the dominant source of prompt photon production at low transverse momenta~\cite{dEnterria:2012kvo}.
To probe the gluon density more precisely, it is necessary to minimise fragmentation photon signals. Not only do they obscure the direct relationship between the outgoing photon and the incoming partons, they also require additional theoretical assumptions to describe their production~\cite{jetphox}, i.e.~fragmentation functions or hadronisation models.
This suppression is accomplished by applying an upper limit on the energy of particles produced in the vicinity of the photon and the resulting observable is denoted ``isolated photon''.
Experiments commonly employ a fixed-cone isolation~\cite{dEnterria:2012kvo}, where the summed energy in a cone of radius $R$ around the photon is required to be below a given threshold.
The same isolation criteria are implemented in both theoretical calculations and experiment.
In addition, requiring prompt photons to be isolated allows suppressing the substantial decay photon background (primarily from $\pi^0\rightarrow\gamma\gamma$ decays) in hadronic collisions.

The unprecedented collision energies provided by the LHC enable the study of prompt photon production at higher $Q^2$ and lower Bjorken-$x$ than was previously accessible.
Isolated prompt photon production has been measured at the LHC in pp~\cite{isophotonsATLAS7TeV1,isophotonsATLAS7TeV2,isophotonsATLAS7TeV3,isophotonsCMS7TeV,isophotonsCMS7TeV2,isophotonsATLAS8TeV,isophotonsATLAS13TeV,isophotonsATLAS13TeVRatios,isophotonsALICE7TeV,isophotonsALICEpp13TeV,isophotonsCMS276TeVppPbPb,isophotonsppandPbPb5TeV}, p--Pb~\cite{isophotonsATLAS8TeVpPb}, and Pb--Pb~\cite{isophotonsCMS276TeVppPbPb,isophotonsATLAS276TeVPbPb,isophotonsppandPbPb5TeV} collisions, with the cross sections well described by perturbative QCD (pQCD) calculations at next-to-leading order (NLO) or next-to-next-to-leading order (NNLO).
While a suppression of the prompt photon cross section at low Bjorken-$x$ is expected due to gluon shadowing in nuclear environments with respect to pp collisions, such suppression has not been observed within the experimental uncertainties in current data due to their limited low-$x$ reach and precision~\cite{isophotonsATLAS8TeVpPb,isophotonsCMS276TeVppPbPb,isophotonsATLAS276TeVPbPb,isophotonsppandPbPb5TeV}.
This consideration motivates the measurement of prompt photon production in p--Pb collisions at lower transverse momentum \pt (lower $x$) than what was previously accessible.

This paper presents new measurements of the isolated prompt photon inclusive production cross section in pp and p--Pb collisions with the ALICE detector.
The measurement for p--Pb collisions is carried out at centre-of-mass energies per nucleon pair of $\sqrt{s_{\text{NN}}}=\SI{5.02}{TeV}$ and \SI{8.16}{TeV} in the transverse momentum range of $12<\pt<\SI{60}{GeV}/c$ and $12<\pt<\SI{80}{GeV}/c$, respectively.
~The low-\pt reach at midrapidity probes gluon densities down to $x\approx 2\pt/\sqrt{s}\approx \num{2.9E-3}$.
This extends the low-$x$ reach in Pb nuclei by almost a factor of two relative to previous measurements in p--Pb collisions~\cite{isophotonsATLAS8TeVpPb}, where nuclear shadowing effects are expected to be sizeable (see Ref.~\cite{NuclearPDFOverview} and calculations below).
The measurement for pp collisions is presented at $\sqrt{s}=\SI{8}{TeV}$.
The inclusive production cross section measurement in pp collisions at $\sqrt{s}=\SI{5.02}{TeV}$ has been published by ALICE in Ref.~\cite{isophotonsppandPbPb5TeV}.

This paper is structured as follows: Section~\ref{sec:alicedetector} describes the ALICE detector, followed by a description of the event selection and triggers in Sec.~\ref{sec:eventselection}.
The reconstruction and identification of photons, as well as the isolation requirements, are discussed in Sec.~\ref{sec:photonreconstruction} and~\ref{sec:isolation}, respectively.
Efficiency and purity corrections of the raw isolated prompt photon yields are introduced in Sec.~\ref{sec:efficiency} and~\ref{sec:purity}.
After a detailed discussion of the arising systematic uncertainties in Sec.~\ref{sec:systematics}, the results of this paper are presented in Sec.~\ref{sec:results}, followed by the conclusion in Sec.~\ref{sec:conclusions}.
\section{ALICE detector}
\label{sec:alicedetector}
A detailed description of the ALICE detector and its performance is provided in Refs.~\cite{ALICE:2008ngc} and~\cite{ALICE:2014sbx}.
We focus here only on those detector subsystems relevant to this measurement, i.e. the calorimeter and tracking systems which are essential for photon reconstruction and isolation, and the determination of the interaction vertex.

The Inner Tracking System (ITS)~\cite{ITS} is the subsystem located closest to the interaction point and consists of six layers of silicon detectors. 
The two innermost layers are Silicon Pixel Detectors (SPD) positioned at radial distances of 3.9\,cm and 7.6\,cm from the beam axis, followed by two layers of Silicon Drift Detectors (SDD) at 15.0\,cm and 23.9\,cm, and two layers of Silicon Strip Detectors (SSD) at 38.0\,cm and 43.0\,cm.
The SDD and SSD have a pseudorapidity coverage of $|\eta|<0.9$ and $|\eta|<1.0$, respectively, while the two SPD layers cover $|\eta|<2$ and $|\eta|<1.4$.
The ITS is used in this analysis for the tracking of charged particles and the reconstruction of the primary collision vertex.

The Time Projection Chamber (TPC)~\cite{TPC} is a large cylindrical drift detector with a two-dimensional ($r\varphi$) position readout on the end plates, while the coordinate along the longitudinal direction ($z$) is obtained from the measured drift time. 
The TPC enables reconstruction of charged particles and their identification via specific energy loss ($\dv*{E}{x}$) measurements.
The TPC covers a pseudorapidity range of $|\eta|<0.9$ over the full azimuth and measures up to 159 individual space points per track.
A solenoid magnet surrounding the central barrel detectors of ALICE provides a magnetic field of $B=\SI{0.5}{T}$, allowing to reconstruct tracks down to $\pt\approx\SI{100}{MeV}/c$.
Combining the tracking capabilities of the ITS and TPC, a transverse momentum resolution of about \SI{1}{\percent} is achieved for $\pt\sim\SI{1}{GeV}/c$, which decreases to about \SI{3}{\percent} at $\SI{10}{GeV}/c$~\cite{TPC}.
For pp collisions at $\sqrt{s}=\SI{8}{TeV}$ and p--Pb collisions at $\sqrt{s_{\text{NN}}}=\SI{8.16}{TeV}$, the TPC and ITS are used to reconstruct charged particles required for photon isolation. 
In p--Pb collisions at $\sqrt{s_{\text{NN}}}=\SI{5.02}{TeV}$, only the ITS is used, resulting in reduced momentum resolution, which however is sufficient for photon identification and isolation~\cite{ALICEGammaHadron}.
This approach was taken to ensure consistency with the detector configuration used in the pp reference at $\sqrt{s}=\SI{5.02}{TeV}$, where for part of the data taking the TPC was excluded from readout to enhance the sampled luminosity~\cite{ALICEGammaHadron}.

The Electromagnetic Calorimeter (EMCal)~\cite{EMCalTDR,EMCalPerf} is a Pb--scintillator sampling calorimeter composed of an alternating stack of $76$ lead absorber and $77$ scintillation layers.
Scintillation light is collected using wavelength-shifting fibres and transported to Avalanche Photodiodes (APDs) for amplification of the scintillation light signal.
The EMCal is located \SI{4.5}{m} in radial distance from the interaction point and covers $|\eta|<0.7$ in pseudorapidity and $80^{\circ} < \varphi < 187^{\circ}$ in azimuthal angle.
It consists of \num{12288} individual towers in total, each with a size of $6\times \SI{6}{cm}^2$ corresponding to roughly two times the Molière radius.
Since 2015, an extension of the EMCal located on the opposite to the EMCal in azimuth, referred to as the Dijet Calorimeter (DCal)~\cite{DCal}, provides additional calorimetric data.
The DCal covers $0.22<|\eta|<0.7$ for $260^{\circ}<\varphi<320^{\circ}$ and $|\eta|<0.7$ for $320^{\circ}<\varphi<327^{\circ}$ with \num{5376} towers. 
From now on, the full calorimeter, comprised of the previously mentioned EMCal and DCal, will be referred to as EMCal and a distinction between the two regions is specified only where required.
The energy resolution of the EMCal 
is $\sigma_E/E = (1.4 \pm
0.1)\%\oplus (9.5 \pm 0.2)\%/\sqrt{E}\oplus (2.9 \pm 0.9) \%/E$, where the energy $E$ is given in units of GeV~\cite{EMCalPerf}.
In this analysis, the EMCal is used to reconstruct and identify photon candidates.
The EMCal also provides hardware triggers, which are used in this analysis and discussed in Sec.~\ref{sec:eventselection}.

The V0 detector~\cite{V0} consists of two scintillator arrays, V0A and V0C, located on opposite sides of the interaction point at $z=+340$\,cm and $z=-90$\,cm and covering $2.8 <\eta < 5.1$ and $-3.7 <\eta < -1.7$, respectively.
The V0 detector provides the minimum bias trigger and is used in this analysis to identify background events originating from beam--gas interactions and out-of-bunch pileup.

\section{Event selection}
\label{sec:eventselection}
The data used in this analysis were collected by the ALICE experiment in the period 2012 to 2016. 
The data for pp collisions at $\sqrt{s}=\SI{8}{TeV}$ and p--Pb collisions at $\sqrt{s_{\text{NN}}}=\SI{5.02}{TeV}$ were recorded in 2012 and 2013, respectively.
The pp collisions data at $\sqrt{s}=\SI{5.02}{TeV}$ and p--Pb collision data at $\sqrt{s_{\text{NN}}}=\SI{8.16}{TeV}$ were recorded in 2017 and 2016, respectively.

Events considered for analysis satisfy at least a minimum bias (MB) trigger condition, which requires the coincidence of signals in the two V0 scintillation arrays.
The cross section of the MB trigger was determined through van der Meer scans to be $\sigma_{\text{MB}}^{\text{pp}}=\SI{55.8\pm1.2}{mb}$~\cite{lumipp8TeV} for pp collisions at $\sqrt{s}=\SI{8}{TeV}$.
For p--Pb and Pb--p collisions at $\sqrt{s_{\text{NN}}}=\SI{8.16}{TeV}$ the cross sections are $\sigma_{\text{MB}}^{\text{p--Pb}}=\SI{2.09\pm0.04}{b}$ and $\sigma_{\text{MB}}^{\text{Pb-p}}=\SI{2.10\pm0.04}{b}$, respectively~\cite{lumipPb8}, where p--Pb denotes the proton beam travelling towards the V0C detector (negative $z$) and Pb--p denotes the proton travelling in the opposite direction.
A similar cross section is measured for p--Pb collisions at $\sqrt{s_{\text{NN}}}=\SI{5.02}{TeV}$ where $\sigma_{\text{MB}}^{\text{p--Pb}}=\SI{2.10\pm0.06}{b}$~\cite{lumipPb5TeV}.

In addition, two EMCal Level 1 photon hardware triggers (L1-$\gamma$) are used to select events with energy depositions above two configurable thresholds~\cite{EMCalPerf}, which are referred to as L1-$\gamma$-low and L1-$\gamma$-high.
A trigger decision, based on a sliding $4\times 4$ tower window using dedicated Trigger Readout Units (TRUs), is issued approximately $\SI{6.5}{\mu s}$ after the bunch crossing.
The trigger thresholds are chosen differently for each dataset, with values summarised in Table~\ref{tab:dataset}. 
The rejection power of the EMCal triggers is determined using the event-normalised energy spectra of clustered energy deposits in the EMCal towers. 
The trigger rejection factor (RF) is determined by fitting the ratio of the spectra in the triggered sample and minimum bias baseline in the energy range well above the trigger threshold.
These rejection factors are corrected via Monte Carlo (MC) simulations for inefficiencies arising from masked TRUs to determine the true inspected luminosity.

Following the trigger selection, timing information from the V0 detectors and correlations between hit points and track segments reconstructed with the SPD are used to remove beam-induced background and out-of-bunch pileup.
In-bunch pileup is mitigated by requiring the reconstruction of at most one primary collision vertex per event in the SPD.
Furthermore, primary vertices with a displacement of more than \SI{10}{cm} along the beam direction from the nominal collision point are rejected.

The nominal integrated luminosity $\mathcal{L}_{\text{int}}= N_{\text{evt}}\times \rm{RF} / \sigma_{\text{MB}}$ and trigger RF are given in Table~\ref{tab:dataset}.
The prompt photon cross section in p--Pb collisions at $\sqrt{s_{\rm NN}}=\SI{5.02}{TeV}$ is determined using both 
L1-$\gamma$ triggered data samples, which are combined using an inverse-variance weighting.
Statistical independence of the two samples is assured by assigning any event fulfilling both L1-$\gamma$ triggers to the low threshold sample.
Only the high threshold L1-$\gamma$ triggered samples are used for the measurement in pp and p--Pb collisions at $\sqrt{s_{\rm NN}}=8$ and \mbox{\SI{8.16}{TeV}}, respectively.
\begin{table}[t]
    \centering
 \caption{Overview of the pp and p--Pb collision data used in this work. For completeness, the data corresponding to pp collisions at $\sqrt{s}=\SI{5.02}{TeV}$~\cite{isophotonsppandPbPb5TeV} are also given. For each EMCal-triggered sample, the trigger thresholds, rejection factors, and integrated luminosities are specified. The given rejection factors and luminosities are corrected for trigger inefficiencies as outlined in the text, except for the measurement in p--Pb collisions at  $\sqrt{s_{\text{NN}}}=\SI{5.02}{TeV}$ (marked with a '$\dagger$' symbol) where the efficiency correction has instead been applied on the level of prompt photon yields and is not included in the given values.}
    \begin{tabular}{llllccc}
    \toprule
      System & $\sqrt{s_{\rm NN}}$ & Trigger & Threshold& RF  & {Int. Luminosity} & Year \\
        & & & (GeV)  &  & {$\mathcal{L}_{\text{int}}$ ($\text{nb}^{-1}$)} &\\
      \midrule
     pp & \SI{5,02}{TeV}~\cite{isophotonsppandPbPb5TeV} &L1-$\gamma$-low&4&\num{997\pm10}&\num{265\pm7} & 2017 \\
    pp & \SI{8,00}{TeV} &L1-$\gamma$-high&10&\num{16372\pm476}&\num{497\pm18} & 2012\\
      \midrule
      p--Pb& $\SI{5.02}{TeV}^{\dagger}$  &L1-$\gamma$-low&7&$1739\pm56$&$0.76\pm 0.03$ & 2013\\
     &&L1-$\gamma$-high&11&$6917\pm 245$&$6.3\pm0.3$ & 2013\\
      \midrule
            p--Pb & \SI{8.16}{TeV}  &L1-$\gamma$-high&8&\num{1231\pm27}&\num{1.41\pm0.04} & 2016\\
    \bottomrule
    \end{tabular}
    \label{tab:dataset}
\end{table}

Correction factors estimated by MC simulations
utilise simulated hard processes based on the PYTHIA~8.2 event generator~\cite{pythia8} using the 2013 Monash Tune~\cite{Monash2013}. In PYTHIA~8, the signal events ($\gamma$--jet) are modelled through $2\to2$ matrix elements for $\rm{gq}\to\gamma \rm{q}$ and $\rm{q\overline{q}}\to\gamma \rm{g}$ hard scatterings at leading order, followed by the leading-logarithm approximations of the parton shower and hadronisation.
In addition, a photon above a given threshold is required to be produced within the EMCal acceptance.
Background events (jet--jet) are simulated requiring hard scatterings with two jets in the final state.
To simulate \pPb~events, the \pp dijet and $\gamma$--jet events simulated with PYTHIA~8 are embedded into \pPb~inelastic collision events generated by \textsc{DPMJET}~\cite{DPMJet} to reproduce the experimentally measured global \pPb~event properties. 

The transport of the generated particles in the detector material is done using GEANT 3~\cite{GEANT3}. 
Following Ref.~\cite{EMCalPerf},  additional corrections are incorporated into the simulation to mimic the observed cross talk between calorimeter cells.  
For the pp and p--Pb collision data at $\snn =8$ and \SI{8.16}{TeV}, respectively, the simulation also included the trigger response taking into account masked TRU's as described in Refs.~\cite{EMCalPerf,isophotonsALICEpp13TeV,isophotonsppandPbPb5TeV}.
For the p--Pb data at $\sqrt{s_{\text{NN}}}=\SI{5.02}{TeV}$, a data-driven evaluation of the acceptance correction arising from the masked TRU's is applied.
\section{Photon reconstruction and identification}
\label{sec:photonreconstruction}
Electromagnetic showers in the EMCal are reconstructed by combining energy depositions in adjacent EMCal towers (from here on referred to as EMCal cells) into clusters, as described in Ref.~\cite{EMCalPerf}.
The clustering algorithm and the photon selection criteria, outlined below, are consistent with what has been used in previous ALICE measurements of isolated prompt photon production~\cite{isophotonsALICE7TeV,isophotonsALICEpp13TeV,isophotonsppandPbPb5TeV}.
In addition, the DCal is included in the analysis of the p--Pb data at $\sqrt{s_{\rm NN}}=\SI{8.16}{TeV}$.
The clustering algorithm begins with a seed cell with an energy of at least \SI{500}{MeV}, subsequently adding adjacent cells which have an energy of at least \SI{100}{MeV}, where the energy thresholds are chosen to reduce sensitivity to detector electronic noise.
Each cluster is required to  contain at least two cells in order to suppress the contribution of single neutrons hitting the readout electronics, which generate single cell clusters.
These neutron-induced spurious signals~\cite{EMCalPerf} are further suppressed by excluding clusters in which cells adjacent to the leading energy cell in a given cluster do not contribute significantly to the total energy of the cluster.
Background from hadronic decays and shower overlaps is suppressed by requiring that each cluster contains at most two local energy maxima.
In order to remove contributions from out-of-bunch pileup, the difference in time between the main bunch crossing and the detection of signal in the highest-energy cell in the cluster is required to satisfy $|\Delta t| <\SI{20}{ns}$ ($-30<\Delta t<\SI{35}{ns}$) for the analysis at $\sqrt{s_{\rm NN}}=\SI{5.02}{TeV}$ (\SI{8.16}{TeV}).

After these selections, which ensure the quality of the cluster sample, photon clusters are identified using the cluster shape and a charged-particle veto.
The shape of a cluster is quantified using the observable $\sigma^2_{\text{long}}$ which corresponds to the length of the long axis of an ellipse characterising the transverse shower shape~\cite{EMCalPerf}.
It is defined as the square of the larger eigenvalue of the energy distribution in the $\eta$--$\varphi$ plane as
\begin{equation}
\sigma^2_{\text{long}} = (\sigma^{2}_{\varphi\varphi} + \sigma^{2}_{\eta\eta})/2 + \sqrt{(\sigma^{2}_{\varphi\varphi} - \sigma^{2}_{\eta\eta})^2/4 + (\sigma^{2}_{\varphi\eta})^2},
\end{equation}
where $\sigma^{2}_{xz} = \langle xz \rangle - \langle x \rangle\langle z \rangle$ are the covariance matrix elements with $\langle x \rangle = (1/w_{\text{tot}})\sum w_i x_i$ that are weighted over all cells in the cluster in $\eta$ and $\varphi$ directions.
The weights are determined for the $i$-th cell in the cluster as $w_i=\mathrm{max}\left(\log(E_{\mathrm{cell},~i}/E_{\mathrm{cluster}}) - w_{0},0\right)$, where $E_{\text{cell},~i}$ and $E_{\rm cluster}$ are the cell and cluster energy, respectively.
The cut-off in the log-weighting is chosen to be $w_{0}=-4.5$ and cells that contain less than {$e^{-4.5} =$ 1.1$\%$} of the total cluster energy are not considered in the $\sigma^2_{\text{long}}$ calculation~\cite{EMCalPerf}. 

The discrimination of photon and $\pi^0$ showers using $\sigma^2_{\text{long}}$ was studied using a detailed simulation of the EMCal response, including the effect of cross talk between adjacent towers, which was found to affect the shower shape~\cite{EMCalPerf}. 
While single photons predominantly generate narrow showers with $\sigma^2_{\text{long}}\approx 0.25$, high-\pt photons from $\pi^0$ and $\eta$ decays generate more elongated clusters due to cluster merging of the decay photons that can no longer be separated on the EMCal surface.
Simulations show that over \SI{90}{\percent} of $\pi^0\rightarrow\gamma\gamma$ decays will be reconstructed as a single cluster for $p_{\rm T}^{\pi^0}>\SI{10}{GeV}/c$~\cite{EMCalPerf}.
Photon clusters from $\pi^0$ and $\eta$ decays are therefore suppressed by requiring $\sigma^2_{\rm long}<0.3$.
Additionally, a threshold of $\sigma^2_{\rm long}>0.1$ is imposed to suppress residual background contributions from spurious signals, such as neutrons hitting the readout electronics.
Overall, good agreement between data and MC is found for the shower shape distributions~\cite{isophotonsALICEpp13TeV,isophotonsppandPbPb5TeV,ALICEGammaHadron}.

Clusters from charged particles are suppressed by utilising a charged particle veto (CPV).
Charged tracks are extrapolated to the EMCal front surface, taking into account the track curvature due to the magnetic field and interactions in the traversed material.
The distance between each cluster and projected track position is calculated on the EMCal surface, and clusters within $|\Delta\varphi|<\SI{0.05}{rad}$ and $|\Delta\eta|<0.05$ of a track are removed from the analysis.
The ratio of cluster energy  and track momentum is required to be below $1.75$, which suppresses accidental cluster--track matches by removing matches in which the track momentum is significantly lower than the energy of the matched cluster.
No CPV is used in the analysis of p--Pb collisions at $\sqrt{s_{\rm NN}}=\SI{5.02}{TeV}$ to ensure a consistent photon identification procedure with respect to the reference pp dataset, which lacks information from the TPC.
This results in a slight loss of signal purity, which is taken into account in the data-driven purity determination outlined in Sec.~\ref{sec:purity}.

\section{Photon isolation}
\label{sec:isolation}
Prompt photon identification in this analysis uses the isolation requirement that the transverse momentum deposited in the vicinity of the photon is below a specified threshold.
The purpose of an isolation requirement is twofold: suppressing the dominant decay photon background and suppressing the contribution of photons produced in the fragmentation of an outgoing parton.
This analysis uses fixed-cone isolation, in which the isolation variable is defined as the sum of the transverse momenta of charged particles within an angular radius, $R =\sqrt{(\varphi_\gamma - \varphi_{\text{ch}})^{2} +(\eta_\gamma - \eta_{\text{ch}})^{2}  } =0.4$, from the cluster position, where $\varphi_\gamma$ and $\eta_{\gamma}$ denote the azimuthal angle and pseudorapidity of the photon, respectively.
Likewise, $\varphi_{\rm ch}$ and $\eta_{\rm ch}$ denote the position of a given track within the isolation cone. 
The isolation energy does not include that of neutral particles, in order to benefit from the larger acceptance of the TPC with respect to the EMCal and to reduce autocorrelation with $\pi^0$ showers. 
Charged particles are reconstructed in the ITS and TPC in data sets where the TPC data are available.
Each track is required to fulfil a set of track quality requirements.
The use of ITS-only tracking at $\sqrt{s}=\SI{5.02}{TeV}$ was not found to introduce significant bias in the isolation selection criterion~\cite{ALICEGammaHadron,isophotonsppandPbPb5TeV}. 

\begin{figure}[t]
    \centering
\includegraphics[width=0.6\textwidth]{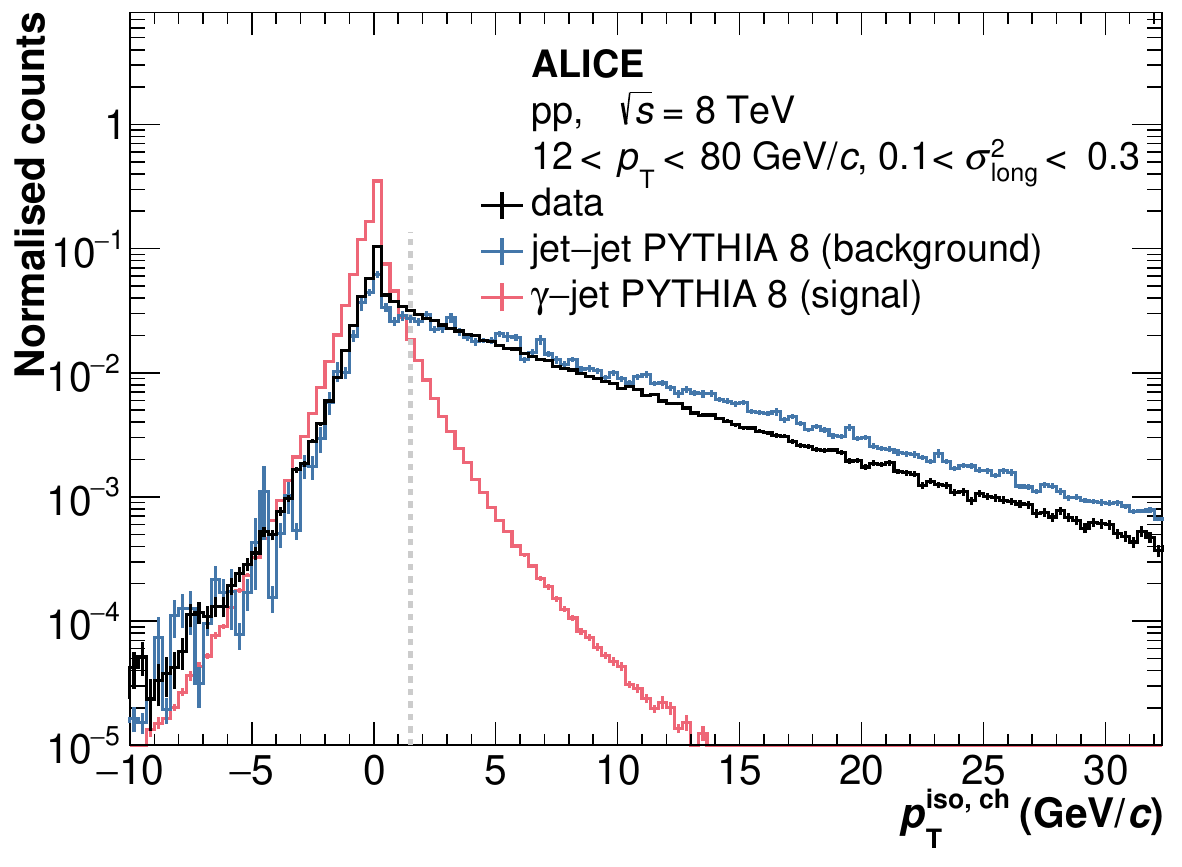}
     \caption{Probability distribution of the charged isolation momentum $p_{\text{\tiny T }}^{\text{iso,~ch}}$ (see Eq.~\ref{eq:isolation}) in pp collisions at \mbox{$\sqrt{s}=\SI{8}{TeV}$}. The isolation threshold $p_{\text{\tiny T }}^{\text{iso,~ch}}<\SI{1.5}{GeV}/c$ is drawn as a dashed grey line. The $p_{\text{\tiny T }}^{\text{iso,~ch}}$ distribution is shown for data (black) and PYTHIA~8 simulations for a signal ($\gamma$--jet) and background (jet--jet) dominated population.}
    \label{fig:isocorrected}
\end{figure}
A candidate cluster is declared isolated when the isolation momentum
\begin{equation}
\pt^\mathrm{iso,~ch} = \sum_{\mathrm{track}~\in~\Delta R<0.4} p_{\mathrm{T}}^{\mathrm{track}} - \rho \times \pi\ \times 0.4^{2},
\label{eq:isolation}
\end{equation}
fulfils $\pt^\mathrm{iso,~ch}<\SI{1.5}{GeV}/c$. 
To account for the soft underlying event (UE) that may produce particles within the isolation cone that do not arise from the hard scattering which generates the EMCal cluster, the charged-particle transverse momentum density $\rho$ is calculated and subtracted in each event. 
Two methods are used to estimate $\rho$. 
The first method, known as the perpendicular cone method, estimates the UE contribution from the energy in two cones of radius $R=0.4$ located $\Delta\varphi=+90^{\circ}$ and $-90^{\circ}$ in azimuth relative to the EMCal cluster, and $\rho$ is the average of the two cone energies divided by their area. 
The second method estimates $\rho$ using the $k_{\rm T}$ jet-finding algorithm with $R=0.4$ from the FASTJET~\cite{fastjet} package, where the median of all reconstructed jet transverse momenta divided by the jet area is used, excluding the two jets with the largest transverse momentum.
The median UE density $\rho$ is less than \SI{1}{GeV}/$c$ and \SI{1.5}{GeV}/$c$ in pp and p--Pb collisions, respectively.
The UE estimation technique was found to have a small impact on the isolated prompt photon cross section, the difference between the two methods is used to assign the corresponding systematic uncertainty, as discussed in Sec.~\ref{sec:systematics}. 

Figure~\ref{fig:isocorrected} shows the distribution of the isolation momentum  $\pt^\text{iso,~ch}$ defined in Eq.~\ref{eq:isolation} for photon candidate clusters in pp collisions at $\sqrt{s}=\SI{8}{TeV}$.
 A requirement of $\pt^\text{iso,~ch}<1.5$\,\GeVc is used to select prompt photon candidates, which results in a signal efficiency of about 90$\%$ and a background rejection of more than \SI{70}{\percent}.
Discrimination of signal and background with this selection is shown using PYTHIA~8 simulations for the signal dominated $\gamma-$jet processes (red) and background photon dominated jet$-$jet processes (blue).
 
 An isolation cone with $R=0.4$ for clusters with $|\eta|>0.5$ does not fit entirely within the acceptance of the tracking detectors, which is $|\eta|<0.9$.
 This is accounted for by a geometrical correction, in which the fraction of the isolation cone that does not fall within the acceptance of the tracking detectors is calculated.
 The magnitude of the correction is at most about \SI{15}{\percent} at the edges of the EMCal acceptance.

\section{Efficiency correction}
\label{sec:efficiency}
The efficiency was estimated using the $\gamma$--jet PYTHIA~8 simulation introduced in Sec.~\ref{sec:eventselection}. 
\begin{figure}[t]
    \centering
    \includegraphics[width=0.6\textwidth]{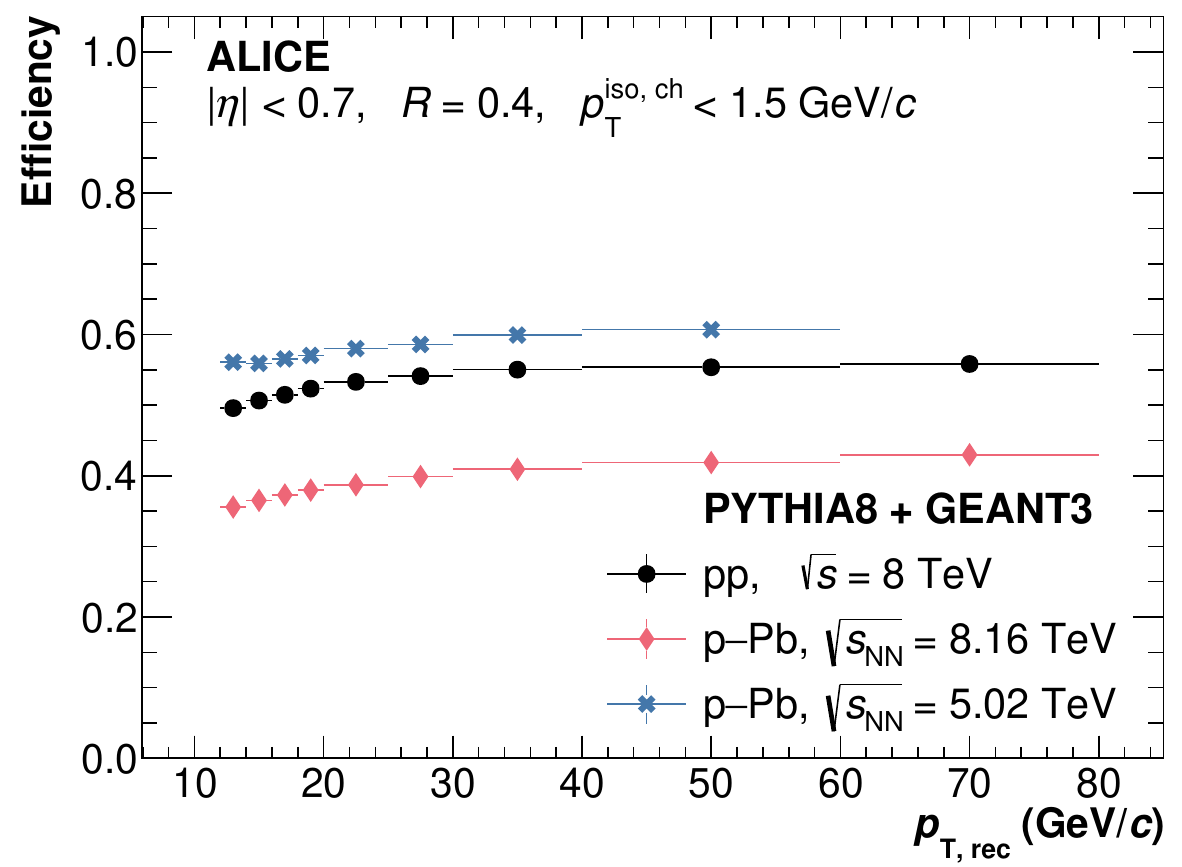}
    \caption{Isolated prompt photon reconstruction efficiencies calculated according to Eq.~\ref{eq:efficiency} using PYTHIA~8 simulations of $\gamma$--jet (signal) processes.}
    \label{fig:efficiencies}
\end{figure}
The reconstruction efficiency is defined as
\begin{equation}
    \epsilon (p_{\text{\tiny T,~rec}}) = \frac{\dv*{N_{\text{clus}}^{\text{sig}}(p_{\text{\tiny T,~rec}})}{p_{\text{\tiny T,~rec}}}}{\dv*{N_{\text{gen}}^{\text{sig}}(p_{\rm T}^\gamma)}{p_{\rm T}^\gamma}},
    \label{eq:efficiency}
\end{equation}
where $N_{\text{clus}}^{\text{sig}}$ is the number of reconstructed clusters that fulfil all the selection criteria and are matched to a signal photon, and $N_{\text{gen}}^{\text{sig}}$ is the number of signal photons at the generator level within the EMCal acceptance.
A signal photon is defined as a photon produced either directly in the hard scattering or in a jet fragmentation that happens to fulfil the charged isolation requirement on generator level with $p_{\rm T}^{\text{iso,~gen}}<\SI{1.5}{GeV}/c$. 
Clusters with a leading contribution from a signal photon are considered true signal clusters.
Bin migration effects due to the finite \pt resolution are included in Eq.~\ref{eq:efficiency} by using the transverse momentum at reconstruction ($p_{\text{\tiny T,~rec}}$) and generator level ($p_{\rm T}^\gamma$) in numerator and denominator, respectively.
Isolation at the generator level $p_{\rm T}^{\text{iso,~gen}}<\SI{1.5}{GeV}/c$ is corrected for UE contributions in order to allow a direct comparison of the results to NLO calculations where no UE is present.
Figure~\ref{fig:efficiencies} shows the isolated prompt photon reconstruction efficiency in pp collisions at $\sqrt{s}=\SI{8}{TeV}$ and p--Pb collisions at $\sqrt{s_{\text{NN}}}=\SI{5.02}{TeV}$ and \SI{8.16}{TeV}.
The efficiency increases slightly with increasing \pt mainly due to increasingly narrow clusters that fulfil the shower shape selection. 
The efficiency is a convolution of the efficiencies of the individual selection criteria.
In the case of pp collisions at $\sqrt{s}=\SI{8}{TeV}$, one e.g. finds an isolation efficiency of more than 90\%, a shower shape selection efficiency between 80\% and 90\%, a track matching efficiency of about 85\% and a cluster reconstruction efficiency of about 70\%.

Differences in the reconstruction efficiencies for the different datasets are attributable to (i) the differing number of masked or dead EMCal channels per run period, (ii) material budget in front of the EMCal, in particular, differences in the number of installed TRD~\cite{TRDPerf} modules from Run 1 to Run 2, (iii) sensitivity of cluster selections to higher multiplicity in p--Pb with respect to pp collisions, and (iv) differences in trigger efficiency, which is sensitive to the number of masked EMCal TRUs.
\section{Purity correction}
\label{sec:purity}
The population of isolated prompt photon candidates still contains a significant contribution from background photons, which originate primarily from neutral meson decays. 
To account for this contamination, the purity of the measurement is determined using two data-driven approaches, known as the ABCD and template fit methods.
Both methods utilise information from the shower shape $\sigma^2_{\text{long}}$ and the isolation momentum $p_{\rm T}^{\text{iso,~ch}}$ to determine the purity in a data-driven way, exploiting the fact that prompt photons tend to produce narrow and isolated clusters in the calorimeter.
It is important to point out that the extraction of the purity in a data-driven way is crucial in order to suppress a model dependence of the result, in particular, any dependence on the absolute prompt photon cross section or signal to background ratio.  
The ABCD method has been used in several previous publications~\cite{isophotonsATLAS7TeV1,isophotonsATLAS7TeV2,ALICE7TeV,isophotonsALICEpp13TeV,isophotonsppandPbPb5TeV}, and is the primary method in this analysis for pp and p--Pb collisions at $\sqrt{s_{\rm NN}}=8$ and \SI{8.16}{TeV}, respectively. 
The template fit method used for the purity estimate of the $\sqrt{s_{\rm NN}}=\SI{5.02}{TeV}$  data was used in a previous publication of isolated photon-hadron correlations in pp and p--Pb collisions at $\sqrt{s_{\rm NN}}=\SI{5.02}{TeV}$~\cite{ALICEGammaHadron}. 
\subsection{The ABCD method}
In the ABCD method, the prompt photon purity is determined using the two-dimensional distribution of shower shape $\sigma^2_{\text{long}}$ and the isolation momentum $p_{\rm T}^{\text{iso,~ch}}$, which is divided into one signal and three control regions. 
Following the derivations given in Ref.~\cite{ALICE7TeV}, the purity $P^{\text{raw}}_{\text{ABCD}}$ in the signal region can be defined as 
\begin{equation}
P^{\text{raw}}_{\text{ABCD}} = 1-\bigg(\frac {N_{\rm n}^{\overline{\rm iso}}/N_{\rm n}^{\rm iso}} {N_{\rm w}^{\overline{\rm iso}}/N_{\rm w}^{\rm iso}}\bigg)_{\rm data}
 \label{eq:ABCDRawPurity}
\end{equation}
where the indices ``n'' and ``w''  refer  to narrow clusters ($0.1<\sigma^2_{\text{long}}<0.3$) and wide clusters ($0.4<\sigma^2_{\text{long}}<2.0$), respectively.  The ``iso'' and ``$\overline{\text{iso}}$'' notations refer to  isolated cluster regions ($\pt^\text{iso,~ch}<\SI{1.5}{GeV}/c$) and anti-isolated cluster regions ($\pt^\text{iso,~ch}>\SI{4}{GeV}/c$), respectively. The number of clusters ($N$) in each region is  the sum of signal (S) + background (B). 

The relation in Eq.~\ref{eq:ABCDRawPurity} holds if (i) the contributions of signal to $N_{\rm w}^{\rm iso}$, $N_{\rm w}^{\overline{\rm iso}}$, and $N_{\rm n}^{\overline{\rm iso}}$ are negligible and (ii) the shower shape and isolation momentum are uncorrelated.
Both assumptions are not fully fulfilled, as (i) there is percent-level leakage of signal into the wide isolated cluster region and (ii) narrow clusters are more likely to be isolated than wide clusters.
These effects are corrected using MC simulations with full detector response to calculate the contribution of underlying correlations via
\begin{equation}
    \alpha_{\text{MC}} = \qty(\frac{B_{\rm n}^{\rm iso}/N_{\rm n}^{\overline{\rm iso}}}{N_{\rm w}^{\rm iso}/N_{\rm w}^{\overline{\rm iso}}})_{\text{MC}},
    \label{eq:alphacorrection}
\end{equation}
where $B_{\rm n}^{\rm iso}$ is the true cluster yield from background sources in the signal region (isolated and narrow clusters).
Typical values of $\alpha_{\text{MC}}$ range from about $1.1$ to $1.4$.
The MC calculation contains both jet--jet and $\gamma$--jet events and the respective yields are scaled with the corresponding cross sections.
The fully corrected purity is then calculated as 
\begin{equation}
    P_{\text{ABCD}} =1-\bigg(\frac {N_{\rm n}^{\overline{\rm iso}}/N_{\rm n}^{\rm iso}} {N_{\rm w}^{\overline{\rm iso}}/N_{\rm w}^{\rm iso}}\bigg)_{\rm data} \times \alpha_{\text{MC}}
\end{equation}
Finally, a bias may arise due to differences in the degree of correlation in data and MC.
This bias is evaluated using the double ratio 
\begin{equation}
    f(\sigma^2_{\text{long}}) = \frac{(N^{\text{iso}}/N^{\overline{\rm iso}})_{\text{data}}}{(N^{\text{iso}}/N^{\overline{\rm iso}})_{\text{MC}}},
    \label{eq:residualcorr}
\end{equation}
where $N^{\text{iso}}$ and $N^{\overline{\rm iso}}$ are the number of isolated and anti-isolated clusters, respectively.
This double ratio is studied for $0.4<\sigma^2_{\text{long}}<2$ where signal contributions are expected to be negligible, and extrapolated into the signal region ($0.1<\sigma^2_{\text{long}}<0.3$) using a first-order polynomial.
A slope consistent with zero is observed within the uncertainties, indicating that an accurate description of correlations is achieved by the MC calculation.
The magnitude of residual mismatch is evaluated in Sec.~\ref{sec:systematics} and taken into account as a systematic uncertainty. 
\subsection{Template fit method}
An alternative approach to determining the purity of the signal sample is a two-component template fit, following the procedure outlined in Ref.~\cite{ALICEGammaHadron}.
The template fit is used for the measurement in p--Pb collisions at $\snn=\SI{5.02}{TeV}$, and found to yield compatible results with the purities obtained using the ABCD method described above~\cite{isophotonsALICEppPbSupplement}.
The $\sigma^2_{\text{long}}$ distribution of the isolated cluster sample is fitted using a linear combination of a self-normalised signal contribution, $\mathbb{S}(\sigma^2_{\text{long}})$, determined from $\gamma$--jet simulations, and a background distribution, $\mathbb{B}(\sigma^2_{\text{long}})$, determined from data using an anti-isolated sideband \mbox{($5 < p_{\rm T}^{\text{iso, ch}}< \SI{10}{GeV}/c$)}.
The two templates are combined linearly via
\begin{equation}
N_{\text{tot}}\qty(\sigma^2_{\text{long}}) = N_{\text{sig}}\times \mathbb{S}\qty(\sigma^2_{\text{long}}) + (N_{\text{data}}-N_{\text{sig}})\times\mathbb{B}\qty(\sigma^2_{\text{long}}),
\end{equation}
where $N_{\text{sig}}$ is the number of signal clusters in the population and is the only free parameter of the fit. $N_{\rm data}$ is the total number of measured isolated clusters and serves as an overall normalisation, whereas $N_{\rm tot}$ denotes the sum of signal and background clusters according to the template fit.
The data are fitted using the MINUIT package~\cite{minuit} using $\chi^2$ minimisation and the MIGRAD algorithm for uncertainty estimation.
The purity $P_{\rm tmp}$ in the signal region is then obtained through the following integrals
\begin{equation}
    P_{\rm tmp} = \frac{\int_{0.1}^{0.3}N_{\text{sig}}\times \mathbb{S}(\sigma^2_{\text{long}})\dd{\sigma^2_{\text{long}}}}{\int_{0.1}^{0.3}N_{\text{tot}}(\sigma^2_{\text{long}})\dd{\sigma^2_{\text{long}}}}.
\end{equation}
Underlying correlations between shower shape and isolation momentum, as introduced in the previous section, are corrected using a re-weighting of the background template via
\begin{equation}
    \mathbb{B}\qty(\sigma^2_{\text{long}}) \rightarrow  \mathbb{B}\qty(\sigma^2_{\text{long}})\times \omega(\sigma^2_{\text{long}}) \qq{with} \omega = \qty(\frac{N^{\text{iso}}(\sigma^2_{\text{long}})}{N^{\overline{\text{iso}}}(\sigma^2_{\text{long}})})_{\text{MC}}
\end{equation}
A residual mismatch in the degree of correlation in MC with respect to data is taken into account in the systematic uncertainties, as described in Sec.~\ref{sec:systematics}.

\subsection{Isolated photon purity}
\begin{figure}[t]
    \centering
    \subfigure[p--Pb $\sqrt{s_{\rm NN}}=\SI{5.02}{TeV}$]{\includegraphics[width=0.49\textwidth]{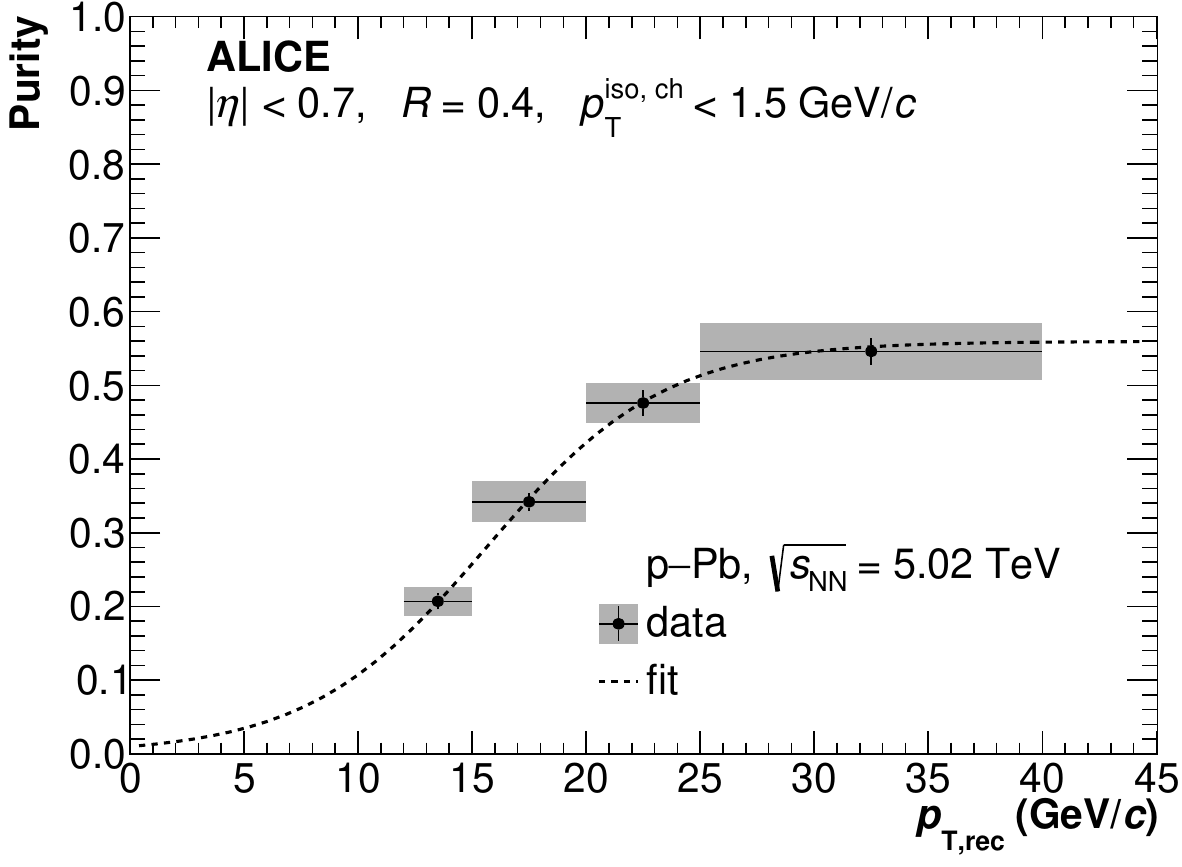}}
    \subfigure[pp $\sqrt{s}=\SI{8}{TeV}$, p--Pb $\sqrt{s_{\rm NN}}=\SI{8.16}{TeV}$]{\includegraphics[width=0.49\textwidth]{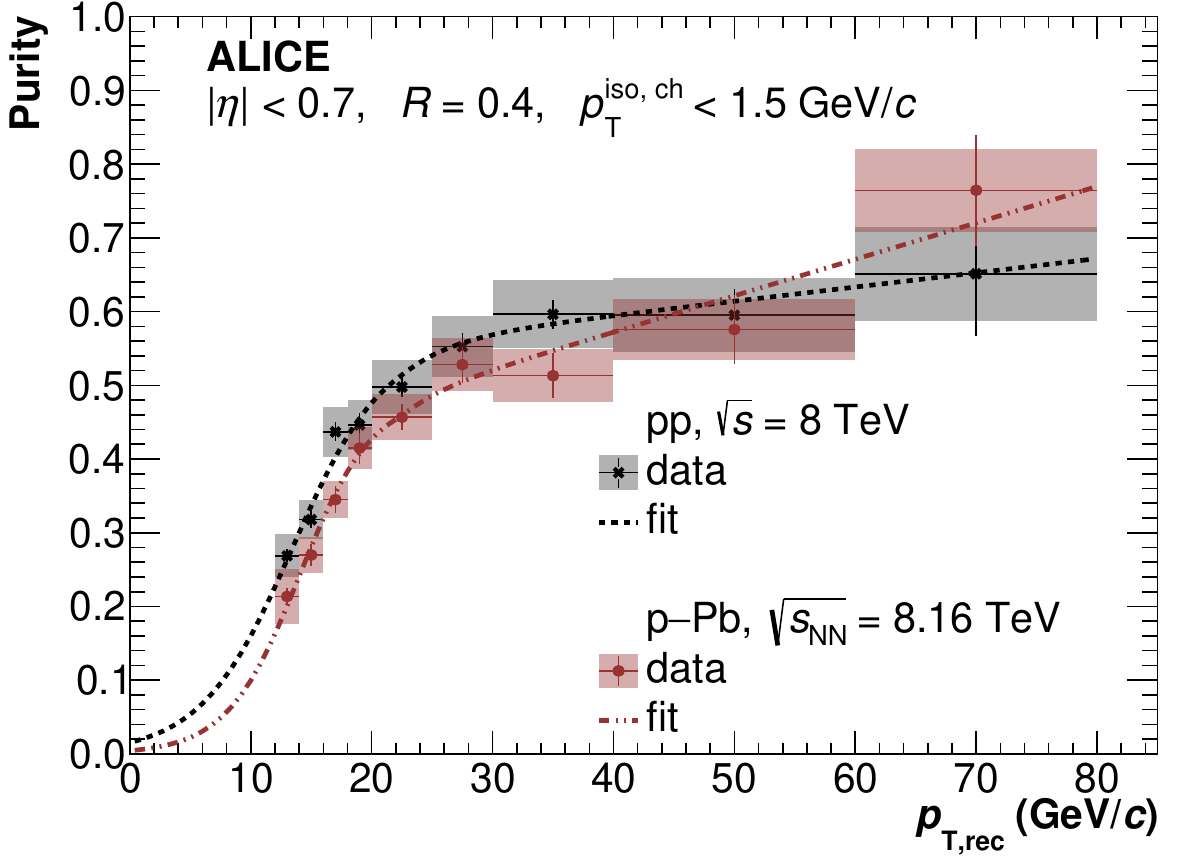}}
    \caption{Isolated prompt photon purity as a function of cluster \pt. The purity in p--Pb collisions at $\snn=\SI{5.02}{TeV}$ (left) is calculated using the template fit approach, whereas for $\snn= 8 $ and \SI{8.16}{TeV} (right) the purity is obtained using the ABCD method. The purity is fitted using a modified sigmoid function given in Eq.~\ref{eq:sigmoid} and an error function for $\snn=\SI{5.02}{TeV}$. Vertical lines and boxes denote statistical and systematic uncertainties, respectively.}
    \label{fig:purities}
\end{figure}

Figure~\ref{fig:purities} shows the purity obtained in pp and p--Pb collisions using the methods described above for the different collision energies.
The signal purity at $\pt\sim\SI{12}{GeV}/c$ is approximately \SI{20}{\percent}, increasing to \SI{70}{\percent} at the highest transverse momentum measured in the analysis.
The shape of the purity distribution is driven by an interplay of physics and detector effects.
An overall increase in purity with increasing \pt is expected from pQCD calculations~\cite{Arleo:2004gn}, where the $\gamma_{\text{prompt}}/\pi^0$ ratio increases. 
The magnitude of this ratio depends on the isolation threshold, which impacts the size of the fragmentation contribution, especially at low-\pt.
The dominant detector effect 
in the prompt photon purity is the finite granularity of the calorimeter and its response to the opening angle of $\pi^0$ and $\eta$ decay photons.
At low \pt the opening angle between the two decay photons is large, resulting in contamination from single decay photons falsely identified as (isolated) prompt photons.
The opening angle decreases with increasing 
meson \pt , resulting in merged showers that are rejected by the shower shape selection and increasing signal purity.
Above $\pt\sim\SI{20}{GeV}/c$ the purity is roughly constant because a large fraction of $\pi^0$ decay clusters merge, leading to contamination in the signal region ($0.1<\sigma^{2}_{\text{long}}<0.3$) that counteracts the increasing $\gamma_{\text{prompt}}/\pi^0$ ratio.
Finally, the data hint at a second rise of the purity for $\pt\gtrsim\SI{40}{GeV}/c$ when almost all merged $\pi^0$ decays produce narrow showers in the signal region and the purity is dominated by the increase of the $\gamma_{\text{prompt}}/\pi^0$ ratio.

The shape of the purity is well described by a sigmoid function including a linear rise at high transverse momentum
\begin{equation}
    P(\pt) = 
     \begin{cases*}
      \frac{a_0}{1+\exp(-a_1 \times (\pt - a_2))} & if $\pt < a_4$ \\
      \frac{a_0}{1+\exp(-a_1 \times (\pt - a_2))} + a_3 (\pt-p_4)      & if $\pt \geq a_4$
    \end{cases*}
    \label{eq:sigmoid}
\end{equation}
where $a_{0...4}$ are  parameters of the fit.
Evaluation of the purity using this parametrisation decreases sensitivity to statistical fluctuations and introduces systematic effects, which  are accounted for in Sec.~\ref{sec:systematics}.
Due to the limited high-\pt reach of the purity estimate for p--Pb collisions at $\snn=5.02$, the fit is performed with $a_3=0$ instead. 
Comparable purities are obtained in p--Pb collisions at $\snn=5.02$ and \SI{8.16}{TeV}, where a slightly higher purity for $\snn=\SI{8.16}{TeV}$ can be attributed to the CPV selections.
Furthermore, a slightly lower purity in p--Pb collisions at $\snn=\SI{8.16}{TeV}$ with respect to pp collisions can be attributed to differences in the datasets outlined in Sec.~\ref{sec:efficiency} as well as the larger underlying event in p--Pb collisions.

While the high-\pt reach of the purity extraction is limited by statistical uncertainties, the low-\pt reach is limited by the fact that a reliable purity extraction becomes increasingly challenging with decreasing \pt.
This is in part due to the low physical signal-to-background ratio as well as the large opening angle of $\pi^0$ decays, which highly reduces the discriminatory power of the shower shape variable at low \pt.
\section{Systematic uncertainties}
\label{sec:systematics}
An overview of the systematic uncertainties of the measurement is given in Table~\ref{tab:syst}.
Systematic uncertainties are specified for the isolated prompt photon cross section in pp and p--Pb collisions across various collision energies, as well as for the nuclear modification factor $R_{\text{pA}}$ (see Sec.~\ref{sec:results}).
The systematic uncertainties are evaluated using variations of the selection criteria employed throughout the analysis.
The variations are chosen to be sufficiently large to adequately sample the resulting changes of the cross section and $R_{\text{pA}}$, which are assumed to follow a Gaussian distribution.
In order to minimise the impact of statistical fluctuations, the resulting deviations are fitted using an appropriate functional form that captures the observed \pt dependence of the respective systematic effect.
Uncertainty sources are grouped into eight categories, where the total uncertainty of a given category is obtained by adding the individual uncertainties in quadrature.
\begin{sidewaystable}
    \centering
    \caption{Systematic uncertainty sources of the isolated prompt photon inclusive production cross section in pp and p--Pb collisions and of the nuclear modification factor $R_{\text{pA}}$. The uncertainties are given for the lowest and highest $p_{\rm T}$ interval of the respective measurement. The total uncertainty is given as the quadratic sum of all uncertainty sources, except for the normalisation uncertainty, which is considered fully \pt correlated and denoted by a separate box in the respective figures.}
    \begin{tabular}{c|cc|cccc|cccc} 
    \toprule
    Name & \multicolumn{2}{c|}{Cross section pp} & \multicolumn{4}{c|}{Cross section p--Pb} & \multicolumn{4}{c}{$R_{\text{pA}}$} \\
     & \multicolumn{2}{c|}{$\SI{8}{TeV}$} & \multicolumn{2}{c}{$\SI{5.02}{TeV}$} & \multicolumn{2}{c|}{$\SI{8.16}{TeV}$} & \multicolumn{2}{c}{$\SI{5.02}{TeV}$} & \multicolumn{2}{c}{$\SI{8.16}{TeV}$}\\
     \midrule
      \pt interval (GeV/$c$)& 12--14 & 60--80 & 12--14 & 40--60 &12--14 & 60--80 & 12--14 & 40--60 & 12--14 & 60--80\\
    \midrule 
    Photon purity & \SI{10.3}{\percent} & \SI{11.5}{\percent} & \SI{12.0}{\percent} & \SI{8.0}{\percent} & \SI{15.2}{\percent} & \SI{11.4}{\percent} & \SI{13.0}{\percent} & \SI{19.0}{\percent} & \SI{8.8}{\percent} & \SI{9.7}{\percent} \\
    Trigger mimicking & \SI{2.0}{\percent} & \SI{2.0}{\percent} & \SI{2.7}{\percent} & \SI{2.7}{\percent} & \SI{4.1}{\percent} & \SI{4.1}{\percent} & \SI{8.4}{\percent} & \SI{8.4}{\percent} & \SI{2.1}{\percent} & \SI{2.1}{\percent} \\
    UE estimation & \SI{2.0}{\percent} & \SI{2.0}{\percent} & \SI{4.0}{\percent} & \SI{4.0}{\percent} & \SI{2.0}{\percent} & \SI{2.0}{\percent} & \SI{4.0}{\percent} & \SI{4.0}{\percent} & \SI{2.0}{\percent} & \SI{2.0}{\percent} \\
    $\sigma_{\rm long}^{2}$ signal range
    & \SI{5.0}{\percent} & \SI{5.0}{\percent} & \SI{1.4}{\percent} & \SI{1.4}{\percent} & \SI{5.0}{\percent} & \SI{5.0}{\percent} & \SI{2.0}{\percent} & \SI{7.5}{\percent} & \SI{4.0}{\percent} & \SI{4.0}{\percent} \\
    CPV & \SI{1.0}{\percent} & \SI{1.0}{\percent} & - & - & \SI{1.0}{\percent} & \SI{1.0}{\percent} & - & - & \SI{1.0}{\percent} & \SI{1.0}{\percent} \\
    Energy scale & \SI{2.0}{\percent} & \SI{2.0}{\percent} & \SI{2.0}{\percent} & \SI{2.0}{\percent} & \SI{2.0}{\percent} & \SI{2.0}{\percent} & $<\SI{0.5}{\percent}$ & $<\SI{0.5}{\percent}$ & $<\SI{0.5}{\percent}$ & $<\SI{0.5}{\percent}$ \\
    Material budget & \SI{2.1}{\percent} & \SI{2.1}{\percent} & \SI{2.1}{\percent} & \SI{2.1}{\percent} & \SI{2.1}{\percent} & \SI{2.1}{\percent} & \SI{3.0}{\percent} & \SI{3.0}{\percent} & \SI{3.0}{\percent} & \SI{3.0}{\percent} \\
    \midrule
    \text{Normalisation}& \multicolumn{2}{c}{$\SI{3.6}{\percent}$} & \multicolumn{2}{c}{$\SI{6.4}{\percent}$} & \multicolumn{2}{c}{$\SI{2.9}{\percent}$} & \multicolumn{2}{c}{$\SI{6.8}{\percent}$} & \multicolumn{2}{c}{$\SI{4.7}{\percent}$}\\
    \midrule\midrule 
    Total  & \SI{12.2}{\percent} & \SI{13.2}{\percent} & \SI{13.3}{\percent} & \SI{11.5}{\percent} & \SI{16.9}{\percent} & \SI{13.6}{\percent} & \SI{16.4}{\percent} & \SI{22.7}{\percent} & \SI{10.7}{\percent} & \SI{11.5}{\percent} \\
    \bottomrule 
    \end{tabular}
    \label{tab:syst}
\end{sidewaystable}
\paragraph{Photon purity}
The  uncertainty associated with the prompt photon purity determination is the dominant source of systematic uncertainty of the measurement, ranging overall from about \SI{8}{\percent} to \SI{15}{\percent} for the prompt photon cross section, where the uncertainty is found to be largest for the lowest purities at low \pt.
The uncertainty is evaluated using variations of the isolation and shower shape selection for the background regions/templates used for the purity estimation. 
In addition, the template fit used for the measurement at $\sqrt{s_{\text{NN}}}=\SI{5.02}{TeV}$ is also performed using only the background template, in order to estimate the uncertainty arising from the signal template obtained from simulations.
The impact of residual correlations between the shower shape and isolation momentum, quantified in Eq.~\ref{eq:residualcorr}, is evaluated by applying an additional MC correction according to the observed residual slopes in the ratio. 
The systematic uncertainty associated with these correlations are largest at low \pt where the purity is low.
The effect of varying the cluster selection criteria as well as the emulation of cross talk between calorimeter cells are evaluated, where the latter was found to be important as it directly affects the shower shape description in MC.
For the ABCD method, the fraction of simulated $\gamma$--jet events with respect to jet--jet events is varied to account for systematic uncertainties arising from the correction denoted in Eq.~\ref{eq:alphacorrection}. 
Variations of the assumed prompt photon spectral shape in the simulations have negligible impact on the applied efficiency and purity corrections~\cite{isophotonsppandPbPb5TeV}.
Finally, the systematic uncertainty associated with  the choice of the functional form to describe the shape of the purity (see Eq.~\ref{eq:sigmoid}) is evaluated through variations of the fit function as well as applying the purity correction point-by-point instead of using any fitting. 
\paragraph{Trigger mimicking}
The efficiency of the EMCal L1 triggers is considered in the measurement using either a full simulation of the TRU response in MC or a data-driven approach using the $\eta$--$\varphi$ distribution of clusters to identify misbehaving TRUs. 
The systematic uncertainty associated with these corrections is estimated conservatively by performing the measurement with and without these corrections and considering half the difference of the fully-corrected cross sections as a systematic uncertainty.
A larger trigger mimicking uncertainty is observed for the $R_{\text{pA}}$ at $\sqrt{s_{{\rm NN}}}=\SI{5.02}{TeV}$ with respect to $\sqrt{s_{{\rm NN}}}=\SI{8.16}{TeV}$, which is mainly driven by the pp reference~\cite{isophotonsALICEppPbSupplement}.
\paragraph{Underlying event estimation}
Systematic uncertainties associated with the underlying event estimation are evaluated by calculating the UE density $\varrho$ using a $k_{\rm T}$ jet finder as well as the perpendicular cone method. 
The difference between the methods on the fully-corrected cross sections is taken as the systematic uncertainty.
\paragraph{Signal shower shape selection and CPV}
Multiple variations of the signal selection according to the shower shape are performed to estimate the associated systematic uncertainty.
Likewise, variations of the $\Delta\varphi$ and $\Delta\eta$ selections used for the CPV are performed to estimate the systematic uncertainty of this procedure, which is found to be about \SI{1}{\percent}.
As no CPV is used for the measurement in p--Pb collisions at $\sqrt{s_{\rm NN}}=\SI{5.02}{TeV}$, no corresponding uncertainty is assigned for this measurement.
\paragraph{Energy scale \& material budget}
The uncertainty on the energy scale of the EMCal is about \SI{0.5}{\percent}~\cite{EMCalPerf,isophotonsppandPbPb5TeV}, which amounts to a \SI{2}{\percent} uncertainty on the steeply falling cross section.
In addition, the material budget uncertainty, which accounts for potential inaccuracies in the MC description of material the photon traverses before reaching the EMCal, has been determined in Ref.~\cite{ALICE:2018mjj} and amounts to \SI{2.1}{\percent}.
\paragraph{Normalisation uncertainty}
Following common convention, the normalisation uncertainty is denoted separately as a small box for the reported cross sections and the $R_{\text{pA}}$.
It accounts for the uncertainty of the minimum-bias cross section $\sigma_{\text{MB}}$ and the uncertainties of the trigger rejection factors.
Both uncertainties are added in quadrature, resulting in a normalisation uncertainty of the integrated luminosity given in Table~\ref{tab:dataset}.

The systematic uncertainties on the $R_{\text{pA}}$ take into account correlations of uncertainties between the p--Pb cross section and the pp reference through a simultaneous variation of selection criteria in both systems.
As the pp reference measurement in pp collisions at $\sqrt{s}=\SI{5.02}{TeV}$ has been reported in Ref.~\cite{isophotonsppandPbPb5TeV} using the ABCD method, we also performed the purity extraction in the same system using the template fitting approach to allow for the cancellation of systematic uncertainties with the measurement in p--Pb collisions at $\sqrt{s_{\text{NN}}}=\SI{5.02}{TeV}$.
Consistency between both purity estimation techniques is demonstrated in Ref.~\cite{isophotonsALICEppPbSupplement}.
Overall, the total systematic uncertainties of the cross sections are less than \SI{18}{\percent} in all systems. 
Partial cancellation of the systematic uncertainties in the p--Pb data with respect to the pp reference is observed, leading to systematic uncertainties of less than \SI{12}{\percent} and \SI{23}{\percent} for the $R_{\text{pA}}$ at $\sqrt{s_{\text{NN}}}=\SI{8.16}{TeV}$ and \SI{5.02}{TeV}, respectively.
Some systematic uncertainty sources, such as e.g. the energy scale uncertainty and material budget uncertainty, are found to be correlated as a function of \pt, amounting to about \SI{65}{\percent} of the systematic uncertainty of the $R_{\rm pA}$.

\section{Results}
\label{sec:results}
\begin{figure}[t]
    \centering
    \subfigure[pp \& p--Pb $\sqrt{s_{\rm NN}}=\SI{5.02}{TeV}$]{\includegraphics[width=0.49\textwidth]{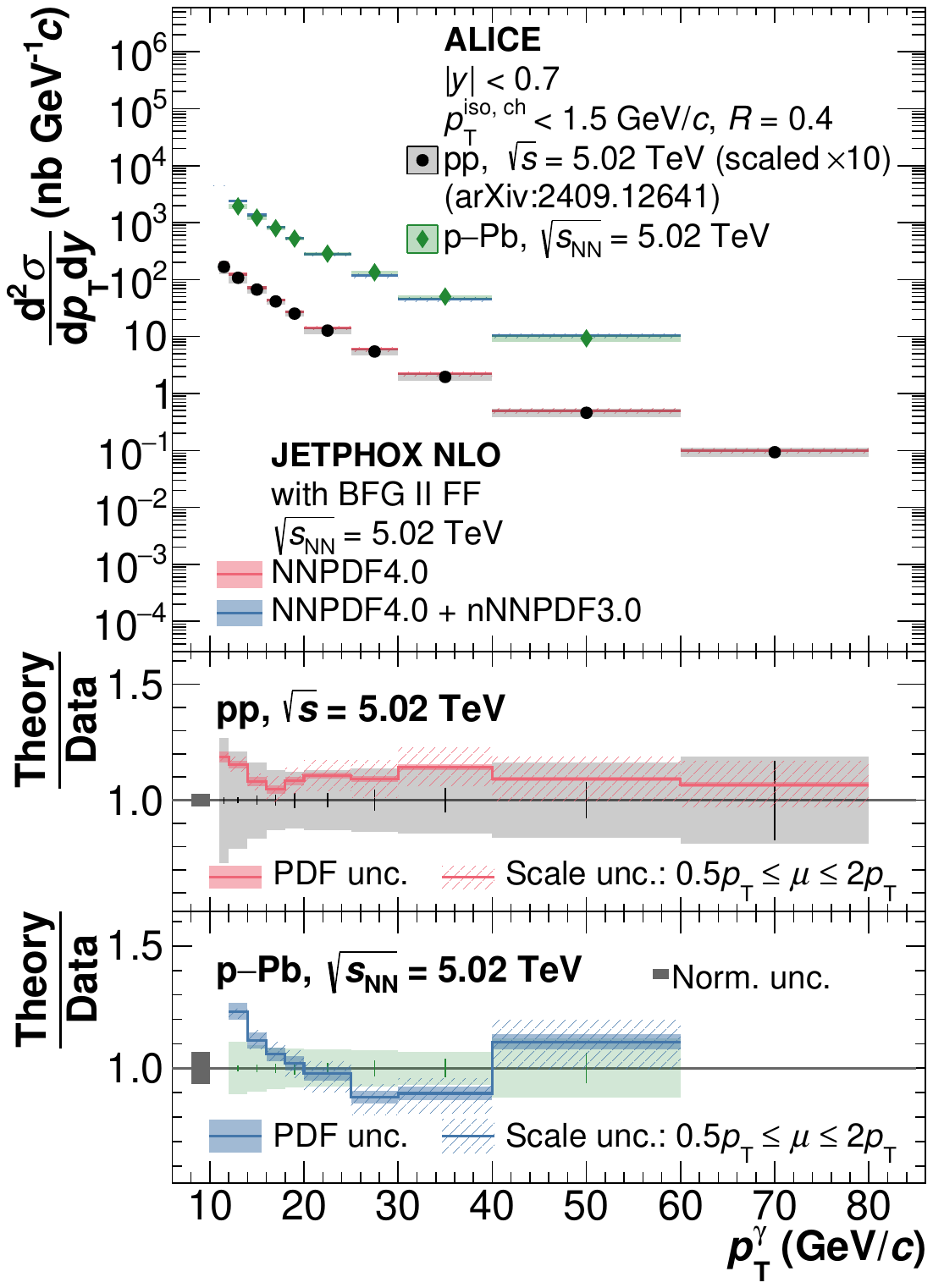}}
    \subfigure[pp $\sqrt{s}=\SI{8}{TeV}$, p--Pb $\sqrt{s_{\rm NN}}=\SI{8.16}{TeV}$]{\includegraphics[width=0.49\textwidth]{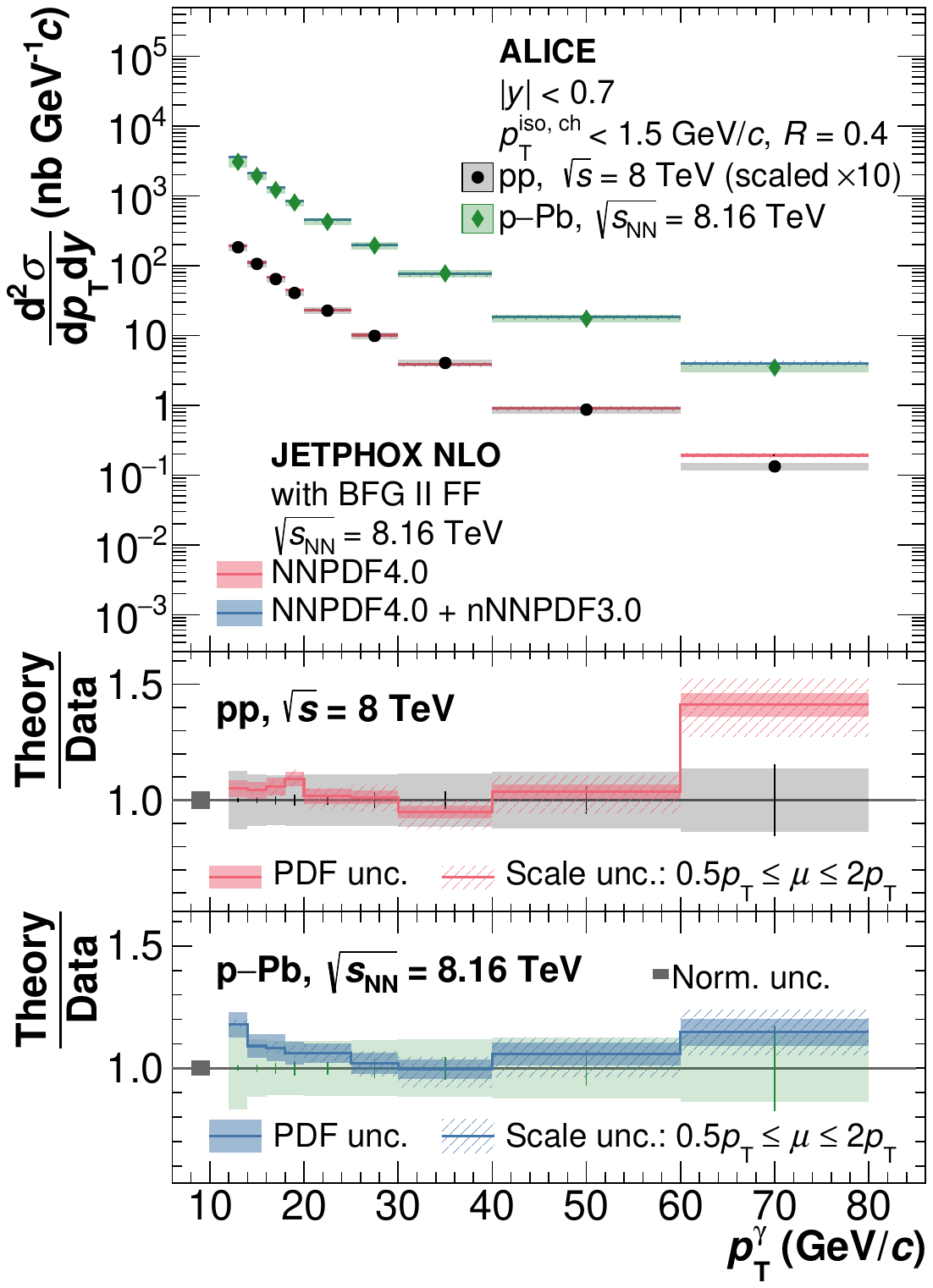}}
    \caption{Isolated prompt photon inclusive production cross section in pp collisions at $\sqrt{s}=5.02$~\cite{isophotonsppandPbPb5TeV} and \SI{8}{TeV}, and p--Pb collisions at $\snn=5.02$ and \SI{8.16}{TeV}. Vertical bars and boxes denote statistical and systematic uncertainties, respectively. 
    Coloured lines denote JETPHOX pQCD calculations at NLO using the recent NNPDF4.0~\cite{NNPDF40} proton PDF and the nNNPDF3.0~\cite{nNNPDF30} nuclear PDF. The BFG II~\cite{BFGII} fragmentation function is used to describe parton-to-photon fragmentation. Dashed coloured bands denote the theoretical scale uncertainties, and solid bands denote PDF uncertainties at 90\% CL.
    \label{fig:crosssections}}
\end{figure}

The isolated prompt photon production cross section is calculated via 
\begin{equation}
    \frac{\dd[2]{\sigma^\gamma}}{\dd{p^{\gamma}_{\rm T}}\dd{y}} = \frac{1}{\mathcal{L}_{\text{int}}}\times \frac{\dd[2]{N_{\rm n}^{\text{iso}}}}{\dd{p_{\rm T}^\gamma}\dd{y}}\times\frac{P}{\epsilon \times \text{Acc}},
\end{equation}
where $N_\text{n}^{\text{iso}}$ is the prompt photon candidate yield, $\mathcal{L_{\text{int}}}$ is the integrated luminosity, $P$ is the purity, $\epsilon$ is the reconstruction efficiency and $\text{Acc}$ is the acceptance of the measurement.
The acceptance is calculated from geometrical considerations and given by the area covered by the calorimeter with respect to the full azimuth at midrapidity ($|y|<0.7$).
Figure~\ref{fig:crosssections} shows the isolated prompt photon inclusive production cross section in pp and p--Pb collisions at $\snn=5.02$, $8$, and \SI{8.16}{TeV}.
The measurement in pp collisions at $\sqrt{s}=\SI{5.02}{TeV}$ was reported in Ref.~\cite{isophotonsppandPbPb5TeV} and is shown for reference.
The cross sections are obtained at midrapidity $(|y|<0.7)$ using a charged isolation momentum threshold $p_{\rm T}^{\text{iso,~ch}}<\SI{1.5}{GeV}/c$ in a cone with radius $R=0.4$.
Vertical bars and boxes denote the statistical and systematic uncertainties, respectively. 
The measurements at both centre-of-mass energies for pp collisions, and for p--Pb collisions at $\sqrt{s_{\rm NN}}=\SI{8.16}{TeV}$, cover a photon transverse momentum range of $12<\pt<\SI{80}{GeV}/c$. 
For p--Pb collisions at $\snn=\SI{5.02}{TeV}$, a slightly lower high-\pt coverage  is reported, due to limitations in the template fit purity determination at high \pt. 

The measurement is compared to pQCD calculations at NLO, which were performed using the JETPHOX~1.3.1~\cite{jetphox} program with an isolation criterion of $p_{\rm T}^{\text{iso}}<\SI{2}{GeV}/c$ in a cone of $R=0.4$.
Since the isolation in JETPHOX corresponds to a limit in final-state radiation within the specified cone, 
a higher isolation threshold than applied on data is chosen to enable comparison with the charged-only isolation employed in the measurements.
The threshold of $\SI{2}{GeV}/c$ has been determined using the neutral energy fraction in the isolation cone in PYTHIA~8 simulations.
The pQCD calculation uses the recent NNPDF4.0~\cite{NNPDF40} proton PDF and the nNNPDF3.0~\cite{nNNPDF30} nuclear PDF to describe the proton and Pb projectiles, respectively.
The (n)PDF uncertainties are determined by performing the calculation for each member of the PDF set and are shown as solid shaded bands.
Prompt photons produced in the fragmentation process are included in the calculation using the BFG II~\cite{BFGII} parton-to-photon fragmentation function. 
The renormalisation scale $\mu_R$, factorisation scale $\mu$, and fragmentation scale $\mu_F$ are chosen to coincide with the photon \pt.
Scale uncertainties are denoted by a dashed band and evaluated through a simultaneous two-point variation of all scales according to $0.5\pt\leq \mu \leq 2\pt$.

The ratios of the pQCD calculations to the respective measurement are shown in the bottom panels of Fig.~\ref{fig:crosssections}.
Agreement is observed for all systems and energies, corroborating previous measurements by the ALICE Collaboration of isolated prompt photon production in pp collisions at $\sqrt{s}=\SI{5.02}{TeV}$~\cite{isophotonsppandPbPb5TeV}, \SI{7}{TeV}~\cite{ALICE7TeV}, and \SI{13}{TeV}~\cite{isophotonsALICEpp13TeV} where likewise agreement between data and theoretical predictions was observed.
Previous findings by the ATLAS Collaboration on prompt photon production in pp and p--Pb collisions at $\snn=\SI{8}{TeV}$ and \SI{8.16}{TeV}~\cite{ATLASpPb8TeV}, respectively, show an underestimation of the data by pQCD calculations at NLO of $10\%$--$15\%$ for $\pt>\SI{25}{GeV}/c$, indicating the need for higher order corrections to accurately describe the data.
While no such discrepancies are observed in the presented measurements within the uncertainties, we speculate that this difference might be due to the less strict isolation criterion  of $E_{\rm T}<\SI{4.8}{GeV} + 4.2 \times 10^{-3}E_{\rm T}^\gamma$ in $R=0.4$ employed in the ATLAS analysis.
This results in a larger contribution from fragmentation photons to the isolated prompt photon cross section, and increases the dependence of the calculations on poorly-constrained parton-to-photon fragmentation functions.

The nuclear modification factor $R_{\text{pA}}$ is defined as
\begin{equation}
    R_{\text{pA}} = \frac{\dd[2]\sigma_{\text{pA}}^\gamma/\dd{\pt}\dd{y^*}}{A_{\text{Pb}}\times\dd[2]\sigma_{\text{pp}}^\gamma/\dd{\pt}\dd{y^*}},
\end{equation}
where $\dd[2]\sigma^\gamma/\dd{\pt}\dd{y^*}$ is the double-differential isolated prompt photon production cross sections in pp and p--Pb collisions, evaluated at rapidity $y^{*}$ in the nucleon--nucleon centre-of-mass frame and $A_{\text{Pb}}=208$ is the mass number of lead.
Differences in collision energy and the rapidity boost $\Delta y =0.46$ due to asymmetric p--Pb collisions in the LHC are taken into account through a scaling of the pp reference.
The scaling is obtained from pQCD calculations at NLO using the JETPHOX program, taking into account both the difference in collision energy and the rapidity boost.
The corresponding correction is less than \SI{2.4}{\percent} in the measured transverse momentum range.

\begin{figure}[t!]
    \centering
   \subfigure[\label{RpAminimalcomparison}]{\includegraphics[width=0.49\textwidth]{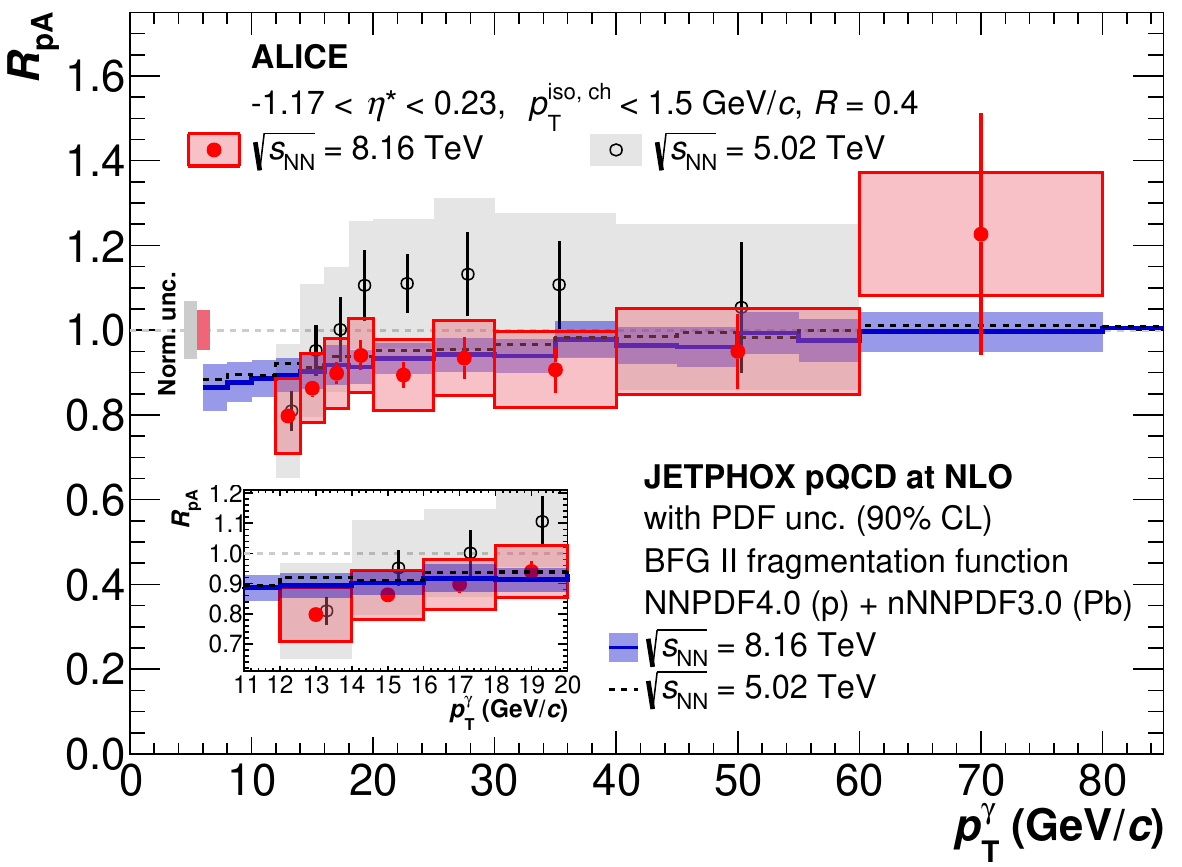}}
    \subfigure[\label{fig:RpATheory}]{\includegraphics[width=0.49\textwidth]{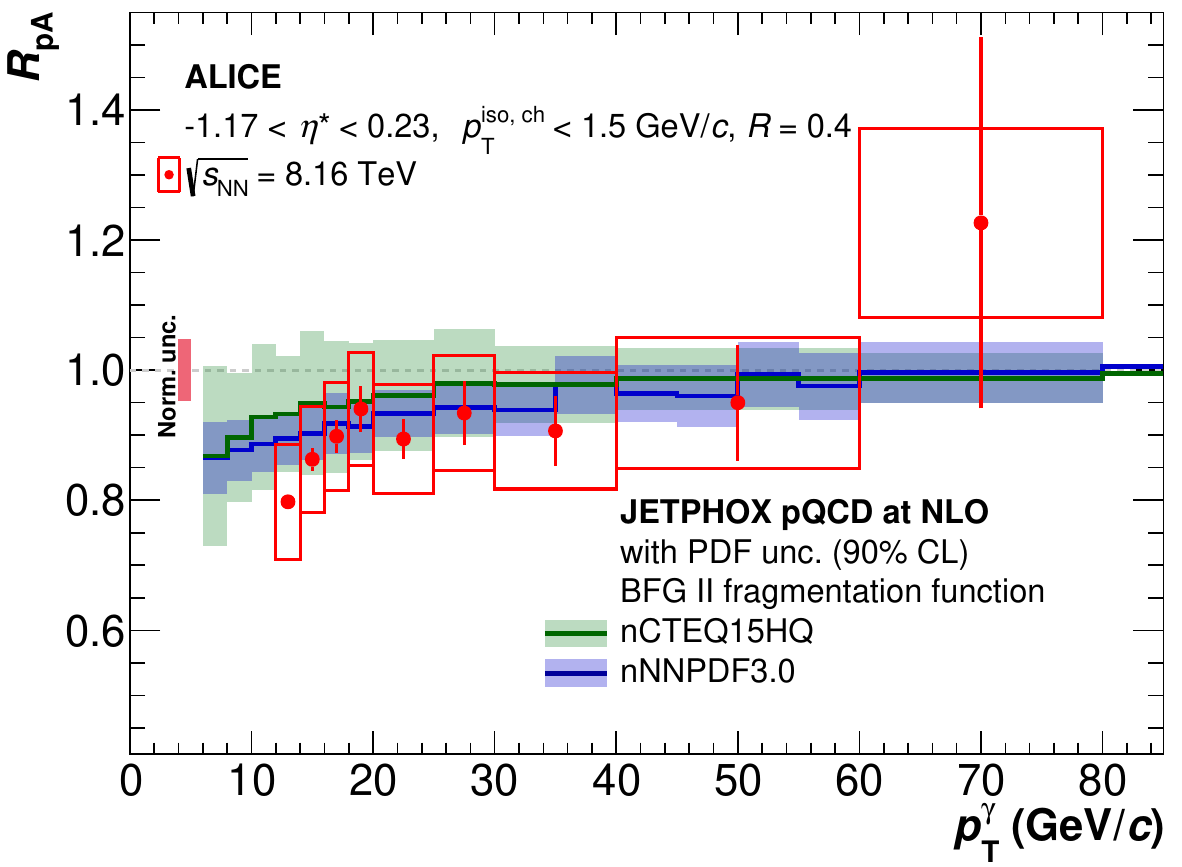}}
    \caption{Nuclear modification factor ($R_{\text{pA}}$) of isolated prompt photon production in p--Pb collisions at $\snn=5.02$ and \SI{8.16}{TeV}. 
    For illustration purposes, the data points at $\sqrt{s_{\text{NN}}}=\SI{5.02}{TeV}$ are displaced by $\Delta p_{\rm T}=+\SI{300}{MeV}/c$. Vertical bars and boxes denote the statistical and systematic uncertainties, respectively. Coloured boxes around unity denote the respective normalisation uncertainties. The measurement is compared to pQCD calculations using recent (n)PDFs, where the shaded band denotes the nPDF uncertainties. The nPDF uncertainties of the prediction at $\snn=5.02$ coincide with those at \SI{8.16}{TeV} and are therefore omitted for visibility. Theoretical scale uncertainties are fully correlated between both collision systems and are therefore not shown. For improved visibility, the right panel shows only the $R_{\text{pA}}$ at $\sqrt{s_{\text{NN}}}=\SI{8.16}{TeV}$ compared to pQCD calculations at NLO using  the nNNPDF3.0~\cite{nNNPDF30} and nCTEQ15HQ~\cite{nCTEQ15HQ} nPDFs.}
    \label{fig:rpa}
\end{figure}

Figure~\ref{fig:rpa} shows the nuclear modification factor $R_{\text{pA}}$ for isolated prompt photon inclusive production at $\snn=\SI{5.02}{TeV}$ and \SI{8.16}{TeV}.
The vertical error bars and boxes denote the statistical and systematic uncertainties, respectively.
As outlined in Sec.~\ref{sec:systematics} and Ref.~\cite{isophotonsALICEppPbSupplement}, the pp reference for the $R_{\rm pA}$ at $\snn=\SI{5.02}{TeV}$ is obtained using the template fit approach to allow for a cancellation of systematic uncertainties.
The statistical uncertainties of the measurement in p--Pb collisions and the pp reference are uncorrelated and therefore added in quadrature.
Systematic uncertainties between the systems are partially correlated, with consequent partial cancellation of uncertainties as outlined in Sec.~\ref{sec:systematics}.
The measured values of $R_{\text{pA}}$ at the two collision energies are in agreement with each other within the uncertainties and consistent with unity for $\pt>\SI{20}{GeV}$.
For $\pt<\SI{20}{GeV}/c$, hints of suppression of the isolated prompt photon production cross section by up to \SI{20}{\percent} are visible, indicating the influence of nuclear effects in the initial state of the collision.
These measurements probe parton densities down to $x_{1,2}\approx 2\pt/\sqrt{s}\approx \num{2.9E-3}$, extending the low-$x$ reach of previous prompt photon measurements in p--Pb collisions~\cite{ATLASpPb8TeV} by about a factor of two.
The significance of the suppression and downward trend for $\pt<\SI{20}{GeV}/c$ are $1.8\sigma$ and $2.3\sigma$ for $\sqrt{s_{\rm NN}}=\SI{8.16}{TeV}$, respectively, taking into account statistical and systematic uncertainties, and their correlation as a function of $p_{\rm T}$.
In particular, we define a downward slope as the deviation from zero of the slope parameter of a first order polynomial.
A suppression $R_{\rm pA}<1$ is quantified using a constant fit and its deviation from unity.
The significance of $R_{\rm pA}<1$ at the lower collision energy of $\sqrt{s_{\rm NN}}=\SI{5.02}{TeV}$ is found to be $1.1\sigma$ for $p_{\rm T}<\SI{14}{GeV}/c$.

\begin{figure}[t]
    \centering
\includegraphics[width=0.8\textwidth]{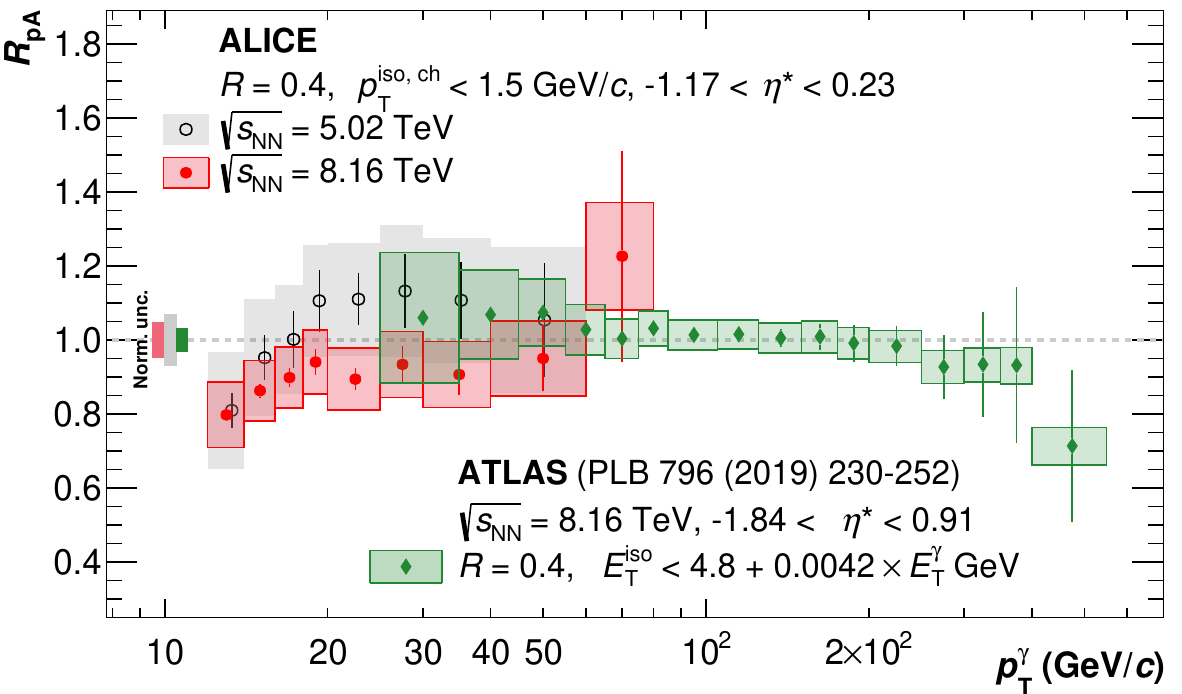}
    \caption{Nuclear modification factor ($R_{\text{pA}}$) of isolated prompt photon production in p--Pb collisions at $\snn=5.02$ and \SI{8.16}{TeV} shown together with a measurement by the ATLAS collaboration in p--Pb collisions at $\snn=\SI{8.16}{TeV}$~\cite{ATLASpPb8TeV}.  For illustration purposes, the data points at $\sqrt{s_{\text{NN}}}=\SI{5.02}{TeV}$ are displaced by $\Delta p_{\rm T}=+\SI{300}{MeV}/c$. The normalisation uncertainties are denoted as coloured boxes around unity.}
    \label{fig:RpAlogx}
\end{figure}
The measured nuclear modification factor is compared to pQCD calculations at NLO  calculated using JETPHOX with the nNNPDF30~\cite{nNNPDF30} and nCTEQ15HQ~\cite{nCTEQ15HQ} nPDFs.
Predictions using nCTEQ15HQ are only shown in Fig.~\ref{fig:RpATheory} for $\snn = \SI{8.16}{TeV}$.
The nPDF uncertainties are denoted by a shaded band.
The scale uncertainties are estimated by simultaneously varying the scales in pp and p--Pb collisions, and were found to fully cancel on the level of the $R_{\text{pA}}$.
Calculations using both nPDFs likewise indicate an increasing suppression of the $R_{\text{pA}}$ with decreasing \pt, which can be attributed to gluon shadowing in the lead nucleus.
Good agreement between the measurement and theoretical calculations is observed within the uncertainties.

A comparison to the $R_{\text{pA}}$ measured at $\snn=\SI{8.16}{TeV}$ by the ATLAS experiment~\cite{ATLASpPb8TeV} is shown in Fig.~\ref{fig:RpAlogx}.
The normalisation uncertainty of the ATLAS data is given in Ref.~\cite{ATLASpPb8TeV} as \SI{2.4}{\percent} for p--Pb and \SI{1.9}{\percent} in pp, which are added in quadrature to obtain the normalisation uncertainty of the $R_{\text{pA}}$.
Agreement between all measurements in the overlapping transverse momentum interval of $25<\pt<\SI{80}{GeV}/c$ is observed.
Furthermore, the figure demonstrates the capabilities of ALICE to measure low-\pt photons with good precision, due to its small material budget, which extends the previously accessible range at low \pt from $\SI{25}{GeV}/c$ to $\SI{12}{GeV}/c$.
\section{Conclusions}
\label{sec:conclusions}
The isolated prompt photon production cross section is reported for pp collisions at $\sqrt{s} = \SI{8}{TeV}$ and p--Pb collisions at $\sqrt{s_{\text{NN}}}= 5.02$ and \SI{8.16}{TeV} by the ALICE experiment.
The measurements are performed at midrapidity ($|y|<0.7$) using an isolation criterion on charged particle transverse momentum $p_{\rm T}^{\text{iso,~ch}}<\SI{1.5}{GeV}/c$ in a cone with radius $R=0.4$ and cover a transverse momentum range from 12 up to $\SI{60}{GeV}/c$ ($\sqrt{s_{\text{NN}}}= \SI{5.02}{TeV}$) and $\SI{80}{GeV}/c$ ($\sqrt{s_{\text{NN}}}= \SI{8.16}{TeV}$).
These measurements probe gluon densities down to $x_{1,2}\approx 2\pt/\sqrt{s}\approx \num{2.9E-3}$, extending the low-$x$ reach of previous prompt photon measurements in p--Pb collisions by about a factor of two.
Good agreement of the data with pQCD calculations at NLO is observed, demonstrating the ability of theoretical calculations using recent (n)PDFs to describe isolated prompt photon production in both collision systems.
The nuclear modification factor is also presented, which agrees with unity and with measurements by ATLAS for $\pt>\SI{20}{GeV}/c$.
At lower transverse momenta, a hint of suppression is observed in the isolated prompt photon cross section in nuclear environments, of up to \SI{20}{\percent} at $p_{\rm T}\sim\SI{12}{GeV}/c$.
At $\sqrt{s_{\text{NN}}}= \SI{8.16}{TeV}$, the suppression increases with decreasing $p_{\rm T}$ with a significance of $2.3\sigma$ for a non-zero slope, and with a significance of $1.8\sigma$ for $R_{\rm pA}<1$.
A suppression with a significance of $1.1\sigma$ is observed for $\sqrt{s_{\text{NN}}}= \SI{5.02}{TeV}$ at $\pt<\SI{14}{GeV}/c$.
These findings are compatible with pQCD calculations at NLO within uncertainties.
The low-$x$ reach of the measurement and the demonstrated sensitivity of prompt photons to nuclear effects offers promising possibilities for future constraints on nuclear PDF fits, where isolated prompt photons provide an important independent probe of the shadowing regime.


\newenvironment{acknowledgement}{\relax}{\relax}
\begin{acknowledgement}
\section*{Acknowledgements}
\input{fa_2025-02-06_Opt_C.tex}
\end{acknowledgement}

\bibliographystyle{utphys}   
\bibliography{bibliography}

\newpage
\appendix

%
%

\section{The ALICE Collaboration}
\label{app:collab}
\input{Alice_Authorlist_2025-02-06_Opt_C.tex}  
\end{document}

%% file: commands.tex
%

\newcommand{\pp}           {pp\xspace}
\newcommand{\ppbar}        {\mbox{$\mathrm {p\overline{p}}$}\xspace}
\newcommand{\XeXe}         {\mbox{Xe--Xe}\xspace}
\newcommand{\PbPb}         {\mbox{Pb--Pb}\xspace}
\newcommand{\pPb}          {\mbox{p--Pb}\xspace}
\newcommand{\AuAu}         {\mbox{Au--Au}\xspace}
\newcommand{\dAu}          {\mbox{d--Au}\xspace}

\newcommand{\snn}          {\ensuremath{\sqrt{s_{\mathrm{NN}}}}\xspace}
\newcommand{\pt}           {\ensuremath{p_{\rm T}}\xspace}
\newcommand{\mt}           {\ensuremath{m_{\rm T}}\xspace}
\newcommand{\meanpt}       {$\langle p_{\mathrm{T}}\rangle$\xspace}
\newcommand{\ycms}         {\ensuremath{y_{\rm CMS}}\xspace}
\newcommand{\ylab}         {\ensuremath{y_{\rm lab}}\xspace}
\newcommand{\etarange}[1]  {\mbox{$\left | \eta \right |~<~#1$}}
\newcommand{\yrange}[1]    {\mbox{$\left | y \right |~<~#1$}}
\newcommand{\dndy}         {\ensuremath{\mathrm{d}N_\mathrm{ch}/\mathrm{d}y}\xspace}
\newcommand{\dndeta}       {\ensuremath{\mathrm{d}N_\mathrm{ch}/\mathrm{d}\eta}\xspace}
\newcommand{\avdndeta}     {\ensuremath{\langle\dndeta\rangle}\xspace}
\newcommand{\dNdy}         {\ensuremath{\mathrm{d}N_\mathrm{ch}/\mathrm{d}y}\xspace}
\newcommand{\Npart}        {\ensuremath{N_\mathrm{part}}\xspace}
\newcommand{\Ncoll}        {\ensuremath{N_\mathrm{coll}}\xspace}
\newcommand{\dEdx}         {\ensuremath{\textrm{d}E/\textrm{d}x}\xspace}
\newcommand{\RpPb}         {\ensuremath{R_{\rm pPb}}\xspace}

\newcommand{\nineH}        {$\sqrt{s}~=~0.9$~Te\kern-.1emV\xspace}
\newcommand{\seven}        {$\sqrt{s}~=~7$~Te\kern-.1emV\xspace}
\newcommand{\twoH}         {$\sqrt{s}~=~0.2$~Te\kern-.1emV\xspace}
\newcommand{\twosevensix}  {$\sqrt{s}~=~2.76$~Te\kern-.1emV\xspace}
\newcommand{\five}         {$\sqrt{s}~=~5.02$~Te\kern-.1emV\xspace}
\newcommand{\twosevensixnn}{$\sqrt{s_{\mathrm{NN}}}~=~2.76$~Te\kern-.1emV\xspace}
\newcommand{\fivenn}       {$\sqrt{s_{\mathrm{NN}}}~=~5.02$~Te\kern-.1emV\xspace}
\newcommand{\LT}           {L{\'e}vy-Tsallis\xspace}
\newcommand{\GeVc}         {Ge\kern-.1emV/$c$\xspace}
\newcommand{\MeVc}         {Me\kern-.1emV/$c$\xspace}
\newcommand{\GeVmass}      {Ge\kern-.2emV/$c^2$\xspace}
\newcommand{\MeVmass}      {Me\kern-.2emV/$c^2$\xspace}
\newcommand{\lumi}         {\ensuremath{\mathcal{L}}\xspace}

\newcommand{\ITS}          {\rm{ITS}\xspace}
\newcommand{\TOF}          {\rm{TOF}\xspace}
\newcommand{\ZDC}          {\rm{ZDC}\xspace}
\newcommand{\ZDCs}         {\rm{ZDCs}\xspace}
\newcommand{\ZNA}          {\rm{ZNA}\xspace}
\newcommand{\ZNC}          {\rm{ZNC}\xspace}
\newcommand{\SPD}          {\rm{SPD}\xspace}
\newcommand{\SDD}          {\rm{SDD}\xspace}
\newcommand{\SSD}          {\rm{SSD}\xspace}
\newcommand{\TPC}          {\rm{TPC}\xspace}
\newcommand{\TRD}          {\rm{TRD}\xspace}
\newcommand{\VZERO}        {\rm{V0}\xspace}
\newcommand{\VZEROA}       {\rm{V0A}\xspace}
\newcommand{\VZEROC}       {\rm{V0C}\xspace}
\newcommand{\Vdecay} 	   {\ensuremath{V^{0}}\xspace}

\newcommand{\ee}           {\ensuremath{e^{+}e^{-}}} 
\newcommand{\pip}          {\ensuremath{\pi^{+}}\xspace}
\newcommand{\pim}          {\ensuremath{\pi^{-}}\xspace}
\newcommand{\kap}          {\ensuremath{\rm{K}^{+}}\xspace}
\newcommand{\kam}          {\ensuremath{\rm{K}^{-}}\xspace}
\newcommand{\pbar}         {\ensuremath{\rm\overline{p}}\xspace}
\newcommand{\kzero}        {\ensuremath{{\rm K}^{0}_{\rm{S}}}\xspace}
\newcommand{\lmb}          {\ensuremath{\Lambda}\xspace}
\newcommand{\almb}         {\ensuremath{\overline{\Lambda}}\xspace}
\newcommand{\Om}           {\ensuremath{\Omega^-}\xspace}
\newcommand{\Mo}           {\ensuremath{\overline{\Omega}^+}\xspace}
\newcommand{\X}            {\ensuremath{\Xi^-}\xspace}
\newcommand{\Ix}           {\ensuremath{\overline{\Xi}^+}\xspace}
\newcommand{\Xis}          {\ensuremath{\Xi^{\pm}}\xspace}
\newcommand{\Oms}          {\ensuremath{\Omega^{\pm}}\xspace}

%% file: fa_2025-02-06_Opt_C.tex

The ALICE Collaboration would like to thank all its engineers and technicians for their invaluable contributions to the construction of the experiment and the CERN accelerator teams for the outstanding performance of the LHC complex.
The ALICE Collaboration gratefully acknowledges the resources and support provided by all Grid centres and the Worldwide LHC Computing Grid (WLCG) collaboration.
The ALICE Collaboration acknowledges the following funding agencies for their support in building and running the ALICE detector:
A. I. Alikhanyan National Science Laboratory (Yerevan Physics Institute) Foundation (ANSL), State Committee of Science and World Federation of Scientists (WFS), Armenia;
Austrian Academy of Sciences, Austrian Science Fund (FWF): [M 2467-N36] and Nationalstiftung f\"{u}r Forschung, Technologie und Entwicklung, Austria;
Ministry of Communications and High Technologies, National Nuclear Research Center, Azerbaijan;
Conselho Nacional de Desenvolvimento Cient\'{\i}fico e Tecnol\'{o}gico (CNPq), Financiadora de Estudos e Projetos (Finep), Funda\c{c}\~{a}o de Amparo \`{a} Pesquisa do Estado de S\~{a}o Paulo (FAPESP) and Universidade Federal do Rio Grande do Sul (UFRGS), Brazil;
Bulgarian Ministry of Education and Science, within the National Roadmap for Research Infrastructures 2020-2027 (object CERN), Bulgaria;
Ministry of Education of China (MOEC) , Ministry of Science \& Technology of China (MSTC) and National Natural Science Foundation of China (NSFC), China;
Ministry of Science and Education and Croatian Science Foundation, Croatia;
Centro de Aplicaciones Tecnol\'{o}gicas y Desarrollo Nuclear (CEADEN), Cubaenerg\'{\i}a, Cuba;
Ministry of Education, Youth and Sports of the Czech Republic, Czech Republic;
The Danish Council for Independent Research | Natural Sciences, the VILLUM FONDEN and Danish National Research Foundation (DNRF), Denmark;
Helsinki Institute of Physics (HIP), Finland;
Commissariat \`{a} l'Energie Atomique (CEA) and Institut National de Physique Nucl\'{e}aire et de Physique des Particules (IN2P3) and Centre National de la Recherche Scientifique (CNRS), France;
Bundesministerium f\"{u}r Bildung und Forschung (BMBF) and GSI Helmholtzzentrum f\"{u}r Schwerionenforschung GmbH, Germany;
General Secretariat for Research and Technology, Ministry of Education, Research and Religions, Greece;
National Research, Development and Innovation Office, Hungary;
Department of Atomic Energy Government of India (DAE), Department of Science and Technology, Government of India (DST), University Grants Commission, Government of India (UGC) and Council of Scientific and Industrial Research (CSIR), India;
National Research and Innovation Agency - BRIN, Indonesia;
Istituto Nazionale di Fisica Nucleare (INFN), Italy;
Japanese Ministry of Education, Culture, Sports, Science and Technology (MEXT) and Japan Society for the Promotion of Science (JSPS) KAKENHI, Japan;
Consejo Nacional de Ciencia (CONACYT) y Tecnolog\'{i}a, through Fondo de Cooperaci\'{o}n Internacional en Ciencia y Tecnolog\'{i}a (FONCICYT) and Direcci\'{o}n General de Asuntos del Personal Academico (DGAPA), Mexico;
Nederlandse Organisatie voor Wetenschappelijk Onderzoek (NWO), Netherlands;
The Research Council of Norway, Norway;
Pontificia Universidad Cat\'{o}lica del Per\'{u}, Peru;
Ministry of Science and Higher Education, National Science Centre and WUT ID-UB, Poland;
Korea Institute of Science and Technology Information and National Research Foundation of Korea (NRF), Republic of Korea;
Ministry of Education and Scientific Research, Institute of Atomic Physics, Ministry of Research and Innovation and Institute of Atomic Physics and Universitatea Nationala de Stiinta si Tehnologie Politehnica Bucuresti, Romania;
Ministry of Education, Science, Research and Sport of the Slovak Republic, Slovakia;
National Research Foundation of South Africa, South Africa;
Swedish Research Council (VR) and Knut \& Alice Wallenberg Foundation (KAW), Sweden;
European Organization for Nuclear Research, Switzerland;
Suranaree University of Technology (SUT), National Science and Technology Development Agency (NSTDA) and National Science, Research and Innovation Fund (NSRF via PMU-B B05F650021), Thailand;
Turkish Energy, Nuclear and Mineral Research Agency (TENMAK), Turkey;
National Academy of  Sciences of Ukraine, Ukraine;
Science and Technology Facilities Council (STFC), United Kingdom;
National Science Foundation of the United States of America (NSF) and United States Department of Energy, Office of Nuclear Physics (DOE NP), United States of America.
In addition, individual groups or members have received support from:
Czech Science Foundation (grant no. 23-07499S), Czech Republic;
FORTE project, reg.\ no.\ CZ.02.01.01/00/22\_008/0004632, Czech Republic, co-funded by the European Union, Czech Republic;
European Research Council (grant no. 950692), European Union;
Deutsche Forschungs Gemeinschaft (DFG, German Research Foundation) ``Neutrinos and Dark Matter in Astro- and Particle Physics'' (grant no. SFB 1258), Germany;
ICSC - National Research Center for High Performance Computing, Big Data and Quantum Computing and FAIR - Future Artificial Intelligence Research, funded by the NextGenerationEU program (Italy). 

%% file: Alice_Authorlist_2025-02-06_Opt_C.tex
\begin{flushleft} 
\small

S.~Acharya\,\orcidlink{0000-0002-9213-5329}\,$^{\rm 50}$, 
A.~Agarwal$^{\rm 133}$, 
G.~Aglieri Rinella\,\orcidlink{0000-0002-9611-3696}\,$^{\rm 32}$, 
L.~Aglietta\,\orcidlink{0009-0003-0763-6802}\,$^{\rm 24}$, 
M.~Agnello\,\orcidlink{0000-0002-0760-5075}\,$^{\rm 29}$, 
N.~Agrawal\,\orcidlink{0000-0003-0348-9836}\,$^{\rm 25}$, 
Z.~Ahammed\,\orcidlink{0000-0001-5241-7412}\,$^{\rm 133}$, 
S.~Ahmad\,\orcidlink{0000-0003-0497-5705}\,$^{\rm 15}$, 
S.U.~Ahn\,\orcidlink{0000-0001-8847-489X}\,$^{\rm 71}$, 
I.~Ahuja\,\orcidlink{0000-0002-4417-1392}\,$^{\rm 36}$
% ...

\section*{Affiliation Notes}

$^{\rm I}$ Deceased\\
$^{\rm II}$ Also at: Max-Planck-Institut fur Physik, Munich, Germany\\
$^{\rm III}$ Also at: Italian National Agency for New Technologies, Energy and Sustainable Economic Development (ENEA), Bologna, Italy\\
$^{\rm IV}$ Also at: Dipartimento DET del Politecnico di Torino, Turin, Italy\\
$^{\rm V}$ Also at: Department of Applied Physics, Aligarh Muslim University, Aligarh, India\\
$^{\rm VI}$ Also at: Institute of Theoretical Physics, University of Wroclaw, Poland\\
$^{\rm VII}$ Also at: Facultad de Ciencias, Universidad Nacional Aut\'{o}noma de M\'{e}xico, Mexico City, Mexico\\

\section*{Collaboration Institutes}

$^{1}$ A.I. Alikhanyan National Science Laboratory (Yerevan Physics Institute) Foundation, Yerevan, Armenia\\
$^{2}$ AGH University of Krakow, Cracow, Poland\\
$^{3}$ Bogolyubov Institute for Theoretical Physics, National Academy of Sciences of Ukraine, Kiev, Ukraine\\
$^{4}$ Bose Institute, Department of Physics  and Centre for Astroparticle Physics and Space Science (CAPSS), Kolkata, India\\
$^{5}$ California Polytechnic State University, San Luis Obispo, California, United States\\
$^{6}$ Central China Normal University, Wuhan, China\\
$^{7}$ Centro de Aplicaciones Tecnol\'{o}gicas y Desarrollo Nuclear (CEADEN), Havana, Cuba\\
$^{8}$ Centro de Investigaci\'{o}n y de Estudios Avanzados (CINVESTAV), Mexico City and M\'{e}rida, Mexico\\
$^{9}$ Chicago State University, Chicago, Illinois, United States\\
$^{10}$ China Nuclear Data Center, China Institute of Atomic Energy, Beijing, China\\
$^{11}$ China University of Geosciences, Wuhan, China\\
$^{12}$ Chungbuk National University, Cheongju, Republic of Korea\\
$^{13}$ Comenius University Bratislava, Faculty of Mathematics, Physics and Informatics, Bratislava, Slovak Republic\\
$^{14}$ Creighton University, Omaha, Nebraska, United States\\
$^{15}$ Department of Physics, Aligarh Muslim University, Aligarh, India\\
$^{16}$ Department of Physics, Pusan National University, Pusan, Republic of Korea\\
$^{17}$ Department of Physics, Sejong University, Seoul, Republic of Korea\\
$^{18}$ Department of Physics, University of California, Berkeley, California, United States\\
$^{19}$ Department of Physics, University of Oslo, Oslo, Norway\\
$^{20}$ Department of Physics and Technology, University of Bergen, Bergen, Norway\\
$^{21}$ Dipartimento di Fisica, Universit\`{a} di Pavia, Pavia, Italy\\
$^{22}$ Dipartimento di Fisica dell'Universit\`{a} and Sezione INFN, Cagliari, Italy\\
$^{23}$ Dipartimento di Fisica dell'Universit\`{a} and Sezione INFN, Trieste, Italy\\
$^{24}$ Dipartimento di Fisica dell'Universit\`{a} and Sezione INFN, Turin, Italy\\
$^{25}$ Dipartimento di Fisica e Astronomia dell'Universit\`{a} and Sezione INFN, Bologna, Italy\\
$^{26}$ Dipartimento di Fisica e Astronomia dell'Universit\`{a} and Sezione INFN, Catania, Italy\\
$^{27}$ Dipartimento di Fisica e Astronomia dell'Universit\`{a} and Sezione INFN, Padova, Italy\\
$^{28}$ Dipartimento di Fisica `E.R.~Caianiello' dell'Universit\`{a} and Gruppo Collegato INFN, Salerno, Italy\\
$^{29}$ Dipartimento DISAT del Politecnico and Sezione INFN, Turin, Italy\\
$^{30}$ Dipartimento di Scienze MIFT, Universit\`{a} di Messina, Messina, Italy\\
$^{31}$ Dipartimento Interateneo di Fisica `M.~Merlin' and Sezione INFN, Bari, Italy\\
$^{32}$ European Organization for Nuclear Research (CERN), Geneva, Switzerland\\
$^{33}$ Faculty of Electrical Engineering, Mechanical Engineering and Naval Architecture, University of Split, Split, Croatia\\
$^{34}$ Faculty of Nuclear Sciences and Physical Engineering, Czech Technical University in Prague, Prague, Czech Republic\\
$^{35}$ Faculty of Physics, Sofia University, Sofia, Bulgaria\\
$^{36}$ Faculty of Science, P.J.~\v{S}af\'{a}rik University, Ko\v{s}ice, Slovak Republic\\
$^{37}$ Faculty of Technology, Environmental and Social Sciences, Bergen, Norway\\
$^{38}$ Frankfurt Institute for Advanced Studies, Johann Wolfgang Goethe-Universit\"{a}t Frankfurt, Frankfurt, Germany\\
$^{39}$ Fudan University, Shanghai, China\\
$^{40}$ Gangneung-Wonju National University, Gangneung, Republic of Korea\\
$^{41}$ Gauhati University, Department of Physics, Guwahati, India\\
$^{42}$ Helmholtz-Institut f\"{u}r Strahlen- und Kernphysik, Rheinische Friedrich-Wilhelms-Universit\"{a}t Bonn, Bonn, Germany\\
$^{43}$ Helsinki Institute of Physics (HIP), Helsinki, Finland\\
$^{44}$ High Energy Physics Group,  Universidad Aut\'{o}noma de Puebla, Puebla, Mexico\\
$^{45}$ Horia Hulubei National Institute of Physics and Nuclear Engineering, Bucharest, Romania\\
$^{46}$ HUN-REN Wigner Research Centre for Physics, Budapest, Hungary\\
$^{47}$ Indian Institute of Technology Bombay (IIT), Mumbai, India\\
$^{48}$ Indian Institute of Technology Indore, Indore, India\\
$^{49}$ INFN, Laboratori Nazionali di Frascati, Frascati, Italy\\
$^{50}$ INFN, Sezione di Bari, Bari, Italy\\
$^{51}$ INFN, Sezione di Bologna, Bologna, Italy\\
$^{52}$ INFN, Sezione di Cagliari, Cagliari, Italy\\
$^{53}$ INFN, Sezione di Catania, Catania, Italy\\
$^{54}$ INFN, Sezione di Padova, Padova, Italy\\
$^{55}$ INFN, Sezione di Pavia, Pavia, Italy\\
$^{56}$ INFN, Sezione di Torino, Turin, Italy\\
$^{57}$ INFN, Sezione di Trieste, Trieste, Italy\\
$^{58}$ Inha University, Incheon, Republic of Korea\\
$^{59}$ Institute for Gravitational and Subatomic Physics (GRASP), Utrecht University/Nikhef, Utrecht, Netherlands\\
$^{60}$ Institute of Experimental Physics, Slovak Academy of Sciences, Ko\v{s}ice, Slovak Republic\\
$^{61}$ Institute of Physics, Homi Bhabha National Institute, Bhubaneswar, India\\
$^{62}$ Institute of Physics of the Czech Academy of Sciences, Prague, Czech Republic\\
$^{63}$ Institute of Space Science (ISS), Bucharest, Romania\\
$^{64}$ Institut f\"{u}r Kernphysik, Johann Wolfgang Goethe-Universit\"{a}t Frankfurt, Frankfurt, Germany\\
$^{65}$ Instituto de Ciencias Nucleares, Universidad Nacional Aut\'{o}noma de M\'{e}xico, Mexico City, Mexico\\
$^{66}$ Instituto de F\'{i}sica, Universidade Federal do Rio Grande do Sul (UFRGS), Porto Alegre, Brazil\\
$^{67}$ Instituto de F\'{\i}sica, Universidad Nacional Aut\'{o}noma de M\'{e}xico, Mexico City, Mexico\\
$^{68}$ iThemba LABS, National Research Foundation, Somerset West, South Africa\\
$^{69}$ Jeonbuk National University, Jeonju, Republic of Korea\\
$^{70}$ Johann-Wolfgang-Goethe Universit\"{a}t Frankfurt Institut f\"{u}r Informatik, Fachbereich Informatik und Mathematik, Frankfurt, Germany\\
$^{71}$ Korea Institute of Science and Technology Information, Daejeon, Republic of Korea\\
$^{72}$ Laboratoire de Physique Subatomique et de Cosmologie, Universit\'{e} Grenoble-Alpes, CNRS-IN2P3, Grenoble, France\\
$^{73}$ Lawrence Berkeley National Laboratory, Berkeley, California, United States\\
$^{74}$ Lund University Department of Physics, Division of Particle Physics, Lund, Sweden\\
$^{75}$ Nagasaki Institute of Applied Science, Nagasaki, Japan\\
$^{76}$ Nara Women{'}s University (NWU), Nara, Japan\\
$^{77}$ National and Kapodistrian University of Athens, School of Science, Department of Physics , Athens, Greece\\
$^{78}$ National Centre for Nuclear Research, Warsaw, Poland\\
$^{79}$ National Institute of Science Education and Research, Homi Bhabha National Institute, Jatni, India\\
$^{80}$ National Nuclear Research Center, Baku, Azerbaijan\\
$^{81}$ National Research and Innovation Agency - BRIN, Jakarta, Indonesia\\
$^{82}$ Niels Bohr Institute, University of Copenhagen, Copenhagen, Denmark\\
$^{83}$ Nikhef, National institute for subatomic physics, Amsterdam, Netherlands\\
$^{84}$ Nuclear Physics Group, STFC Daresbury Laboratory, Daresbury, United Kingdom\\
$^{85}$ Nuclear Physics Institute of the Czech Academy of Sciences, Husinec-\v{R}e\v{z}, Czech Republic\\
$^{86}$ Oak Ridge National Laboratory, Oak Ridge, Tennessee, United States\\
$^{87}$ Ohio State University, Columbus, Ohio, United States\\
$^{88}$ Physics department, Faculty of science, University of Zagreb, Zagreb, Croatia\\
$^{89}$ Physics Department, Panjab University, Chandigarh, India\\
$^{90}$ Physics Department, University of Jammu, Jammu, India\\
$^{91}$ Physics Program and International Institute for Sustainability with Knotted Chiral Meta Matter (WPI-SKCM$^{2}$), Hiroshima University, Hiroshima, Japan\\
$^{92}$ Physikalisches Institut, Eberhard-Karls-Universit\"{a}t T\"{u}bingen, T\"{u}bingen, Germany\\
$^{93}$ Physikalisches Institut, Ruprecht-Karls-Universit\"{a}t Heidelberg, Heidelberg, Germany\\
$^{94}$ Physik Department, Technische Universit\"{a}t M\"{u}nchen, Munich, Germany\\
$^{95}$ Politecnico di Bari and Sezione INFN, Bari, Italy\\
$^{96}$ Research Division and ExtreMe Matter Institute EMMI, GSI Helmholtzzentrum f\"ur Schwerionenforschung GmbH, Darmstadt, Germany\\
$^{97}$ Saga University, Saga, Japan\\
$^{98}$ Saha Institute of Nuclear Physics, Homi Bhabha National Institute, Kolkata, India\\
$^{99}$ School of Physics and Astronomy, University of Birmingham, Birmingham, United Kingdom\\
$^{100}$ Secci\'{o}n F\'{\i}sica, Departamento de Ciencias, Pontificia Universidad Cat\'{o}lica del Per\'{u}, Lima, Peru\\
$^{101}$ Stefan Meyer Institut f\"{u}r Subatomare Physik (SMI), Vienna, Austria\\
$^{102}$ SUBATECH, IMT Atlantique, Nantes Universit\'{e}, CNRS-IN2P3, Nantes, France\\
$^{103}$ Sungkyunkwan University, Suwon City, Republic of Korea\\
$^{104}$ Suranaree University of Technology, Nakhon Ratchasima, Thailand\\
$^{105}$ Technical University of Ko\v{s}ice, Ko\v{s}ice, Slovak Republic\\
$^{106}$ The Henryk Niewodniczanski Institute of Nuclear Physics, Polish Academy of Sciences, Cracow, Poland\\
$^{107}$ The University of Texas at Austin, Austin, Texas, United States\\
$^{108}$ Universidad Aut\'{o}noma de Sinaloa, Culiac\'{a}n, Mexico\\
$^{109}$ Universidade de S\~{a}o Paulo (USP), S\~{a}o Paulo, Brazil\\
$^{110}$ Universidade Estadual de Campinas (UNICAMP), Campinas, Brazil\\
$^{111}$ Universidade Federal do ABC, Santo Andre, Brazil\\
$^{112}$ Universitatea Nationala de Stiinta si Tehnologie Politehnica Bucuresti, Bucharest, Romania\\
$^{113}$ University of Derby, Derby, United Kingdom\\
$^{114}$ University of Houston, Houston, Texas, United States\\
$^{115}$ University of Jyv\"{a}skyl\"{a}, Jyv\"{a}skyl\"{a}, Finland\\
$^{116}$ University of Kansas, Lawrence, Kansas, United States\\
$^{117}$ University of Liverpool, Liverpool, United Kingdom\\
$^{118}$ University of Science and Technology of China, Hefei, China\\
$^{119}$ University of South-Eastern Norway, Kongsberg, Norway\\
$^{120}$ University of Tennessee, Knoxville, Tennessee, United States\\
$^{121}$ University of the Witwatersrand, Johannesburg, South Africa\\
$^{122}$ University of Tokyo, Tokyo, Japan\\
$^{123}$ University of Tsukuba, Tsukuba, Japan\\
$^{124}$ Universit\"{a}t M\"{u}nster, Institut f\"{u}r Kernphysik, M\"{u}nster, Germany\\
$^{125}$ Universit\'{e} Clermont Auvergne, CNRS/IN2P3, LPC, Clermont-Ferrand, France\\
$^{126}$ Universit\'{e} de Lyon, CNRS/IN2P3, Institut de Physique des 2 Infinis de Lyon, Lyon, France\\
$^{127}$ Universit\'{e} de Strasbourg, CNRS, IPHC UMR 7178, F-67000 Strasbourg, France, Strasbourg, France\\
$^{128}$ Universit\'{e} Paris-Saclay, Centre d'Etudes de Saclay (CEA), IRFU, D\'{e}partment de Physique Nucl\'{e}aire (DPhN), Saclay, France\\
$^{129}$ Universit\'{e}  Paris-Saclay, CNRS/IN2P3, IJCLab, Orsay, France\\
$^{130}$ Universit\`{a} degli Studi di Foggia, Foggia, Italy\\
$^{131}$ Universit\`{a} del Piemonte Orientale, Vercelli, Italy\\
$^{132}$ Universit\`{a} di Brescia, Brescia, Italy\\
$^{133}$ Variable Energy Cyclotron Centre, Homi Bhabha National Institute, Kolkata, India\\
$^{134}$ Warsaw University of Technology, Warsaw, Poland\\
$^{135}$ Wayne State University, Detroit, Michigan, United States\\
$^{136}$ Yale University, New Haven, Connecticut, United States\\
$^{137}$ Yildiz Technical University, Istanbul, Turkey\\
$^{138}$ Yonsei University, Seoul, Republic of Korea\\
$^{139}$ Affiliated with an institute formerly covered by a cooperation agreement with CERN\\
$^{140}$ Affiliated with an international laboratory covered by a cooperation agreement with CERN.\\

\end{flushleft} 